\documentclass[twocolumn,prb]{revtex4}

  \usepackage[dvips]{graphicx}

\usepackage{dcolumn}
\usepackage{amsmath}
\usepackage{latexsym}

\begin{document}

\title[Short Title]{Flux-driven Josephson traveling-wave parametric amplifier}
\author{A. B. Zorin}
\affiliation{Physikalisch-Technische Bundesanstalt, Bundesallee
100, 38116 Braunschweig, Germany}


\begin{abstract}
We have developed a concept for a traveling-wave parametric amplifier
driven by a magnetic-flux wave. The circuit consists of a serial array of symmetric
dc SQUIDs coupled inductively to a separate superconducting $LC$ transmission
line carrying the pump wave. The adjusted phase velocity of the pump flux-wave of frequency $\omega_p$
ensures amplification of the signal ($\omega_s$) and the idler ($\omega_i$) waves, with frequencies
obeying the relation $\omega_s+\omega_i = \omega_p$.
The advantage of the proposed flux-driven linear circuit includes a
large gain in a wide frequency range and overcoming of the pump depletion problem.
Unlike the conventional traveling-wave amplifiers, the signal and pump in the proposed circuit
are applied to different ports, what can greatly simplify the whole measurement setup.
The experimental parameters and characteristics of this amplifier have been evaluated
and show promise for applications in quantum information single-photon circuits.


\end{abstract}
\maketitle

\section{Introduction}

Josephson parametric amplifiers (JPAs) are among the most useful tools
in the field of quantum technology (see, e.g.,
Refs. \cite{Castellanos-Beltran2008,Clerk2010,Abdo2011,Vijay2011,Flurin2012,Devoret2013,Eichler2014,Vool2016,Lecocq2017,Sivak2019}).
However, the quantum-limited performance of these devices is often combined
with a fairly limited bandwidth and small dynamic range. The limitation of the bandwidth is primarily
due to an amplifier cavity that enhances the interaction between the signal and
the pump, thereby ensuring a large parametric gain.
For this reason, JPAs that are based on traveling microwaves (i.e., that have no cavity, and are thus free
from the gain-bandwidth trade-off) have been extensively explored in the past several
years. \cite{Mohebbi2011,Yaakobi2013,OBrien2014,White2015,Macklin2015,Bell-Samolov2015,Zorin2016,Zorin2017,Miano2018,WenyuanZhang2017}
Moreover, the lack of a cavity enables a larger dynamic range in these amplifiers. \cite{OBrien2014}

The traveling-wave JPAs (TWJPAs) have an architecture of a microwave
transmission line made of periodically repeated sections,
including either single Josephson junctions \cite{Mohebbi2011,Yaakobi2013,OBrien2014,White2015,Macklin2015}
or various types of superconducting quantum-interference devices (SQUIDs). \cite{Bell-Samolov2015,Zorin2016,Zorin2017,Miano2018,WenyuanZhang2017}
Most of these amplifiers operate in the four-wave-mixing mode (i.e., when $\omega_s+\omega_i = 2\omega_p$,
where $\omega_s$, $\omega_i$, and $\omega_p$ are the signal, idler and pump frequencies, respectively).
This mode is possible due to the natural Kerr-like nonlinearity of the Josephson
inductance $L_J= d\Phi(I)/dI$.
This inductance, has a quadratic dependence on the reasonably small alternating
current $I(t)$ (see, e.g., Refs. \cite{Yaakobi2013,White2015}):
\begin{eqnarray}
L_J(t) 
\approx \left[1+ 0.5I^2(t)/I_c^2\right]L_{J0},
\label{L-4ph}
\end{eqnarray}
where the parameter $L_{J0} = \varphi_0/I_c$ is the linear
(small-signal) inductance, $\varphi_0 = \hbar/2e$ is the reduced flux quantum, and $I_c$ is the critical
current.

Recently, the concept of a TWJPA with three-wave
mixing was proposed \cite{Zorin2016} and tested. \cite{Zorin2017, Miano2018}
In this amplifier, the frequencies
obey the relation $\omega_s+\omega_i = \omega_p$ and the performance is based on
the noncentrosymmetric nonlinearity of type
$L_J^{-1}(\Phi) = dI(\Phi)/d\Phi \approx (1 - \beta_L \Phi/\varphi_0)/L
\neq L_J^{-1}(-\Phi)$ 
or, equivalently,
\begin{equation}
L_J(t) \approx \left[1 + \beta_L^2 I(t)/I_c\right]L,
\label{L-3ph}
\end{equation}
which can be engineered with the help
of the flux-biased nonhysteretic rf SQUID
having linear geometrical inductance $L$ and dimensionless screening
parameter $\beta_L \equiv LI_c/\varphi_0 < 1$ (or, alternatively,
by using an asymmetric SQUID having a Josephson kinetic inductance in
one branch \cite{Sivak2019,Zorin2017,Frattini2017,Frattini2018}).
Compared with the TWJPA with four-wave mixing exploiting Kerr nonlinearity, this three-wave mixing TWJPA
was almost free of self-modulation and cross-modulation effects. So, the corresponding wave numbers
$k_s$, $k_i$ and $k_p$ did not depend on the pump power $P_p$, which may
otherwise cause a significant phase mismatch $\Delta k = k_p-k_s-k_i \neq 0$ \cite{Agrawal}
and require careful dispersion engineering (resonant phase
matching). \cite{White2015,Macklin2015}

Both these types of TWJPA were based
on Josephson nonlinearity (i.e., the dependence of the Josephson inductance $L_J(I)$
on current $I(t)$), enabling time-variation of $L_J(t)$ with the pump frequency $\omega_p$
or the double pump frequency $2\omega_p$
in the three-wave-mixing and four-wave-mixing regimes, respectively.
The later is the principle of operation of an optical-fiber traveling-wave parametric amplifier \cite{Agrawal}
(as well as a superconducting kinetic-inductance-based traveling-wave
parametric amplifier \cite{Eom2012}), where a large pump ensures modulation of the
refraction index of optical medium (or kinetic inductance of the wire
in the microwave transmission line)
with weak Kerr
nonlinearity and hence enabling a growth of a signal.
Unfortunately, in Josephson parametric amplifiers, the pump power $P_p$ (more
specifically, the pump current $I_p$) is limited
by the critical current of the junctions, $\sqrt{P_p} \propto |I_p| \lesssim 0.5 I_c$.
In addition, the pump power is
progressively depleted due to conversion into the signal power $P_s$ and the idler power $P_i$.
This sets a limitation on the length and therefore on the gain of the TWJPA.
Thus, similarly to the case of cavity-based JPAs, \cite{Kamal2009,Abdo2013}
the pump depletion reduces the compression point and limits
the dynamic range of the TWJPA. \cite{OBrien2014,Zorin2016,Kylemark2006}
Finally, the nonlinearity may cause
unwanted leakage of the pump power by converting it into
power of higher harmonics. \cite{White2015,Zorin2016}

In this paper, we propose an alternative concept for a realization of a TWJPA
in which these problems can be mitigated.
Our TWJPA uses the principle of \emph{straightforward modulation}
of the Josephson element (dc SQUID) by means of an external flux
drive $\Phi_\textrm{ext}(t)$, that is,
\begin{equation}
L_{J}(t) = L_{J0}\left[\Phi_\textrm{ext}(t)\right],
\label{L-direct-var}
\end{equation}
applied via a separate pump line.
In this case, Josephson nonlinearity is no longer used for mixing
the modes, so the parametric circuit can operate in the regime with
time-varying \emph{linear} inductance $L_J(t)$ and hence have a larger
dynamic range.

This fundamental principle of parametric amplification (see, e.g.,
Refs.~\cite{Landau-Lifshitz-1,Migulin}) was realized by
Yamamoto et al. \cite{Yamamoto2008} in a cavity-based JPA by directly
modulating the flux threading the dc SQUID at twice the resonator frequency.
The SQUID in this inherently linear circuit played the role of a flux-dependent inductor
terminating a superconducting-transmission-line $\lambda/4$ resonator.
The great advantage of this JPA was that the pump and the signal ($\omega_s \approx 0.5 \omega_p$)
were applied to different ports of the circuit, so the separation of the output signal from the pump
was straightforward. Lately, these flux-driven cavity-based JPAs were successfully
applied, for example, for microwave photon generation, \cite{Wilson2010,Wilson2011}
readout of a flux qubit, \cite{Lin2013} ultra-sensitive thermal microwave
detectors, \cite{Simbierowicz2018} and squeezing of vacuum
fluctuations. \cite{Zhong2013} The nonlinear effects in flux-driven SQUID-based
circuits were studied in Refs.~\cite{Krantz2013,Boutin2017}.
Another variant of a flux-modulated JPA with high gain was based on eight serially-connected
dc SQUIDs that were embedded in a lumped-element $LC$ circuit and driven in phase
by an rf current in the common coupling loop.  \cite{Zhou2014}

However, the traveling-wave case is unique in that it
requires the flux-drive phases on the individual cells
of the SQUID-based transmission line to be properly adjusted. Only then
does the parametric amplification of a traveling signal become possible.
Below we show that such flux drive is possible with the help of a separate pump transmission line;
because this pump line is inductively (weakly) coupled to the signal line over its full length,
the flux drive can be realized in the desired traveling-wave fashion.
Importantly, in the case of sufficiently small coupling of two lines,
the power transmitted in the isolated pump line is no longer limited
by the critical current of the Josephson junctions (SQUIDs) inserted in the signal line,
but it is limited only by the tolerance of the cryogenic setup
to a generated Joule heat. Therefore, sufficiently
high pump power can ensure an almost-depletion-free pump regime of parametric amplification.
In this case, each elementary cell in the signal line provides the maximum possible gain.

\begin{figure}[t]
\begin{center}
\includegraphics[width=3.5in]{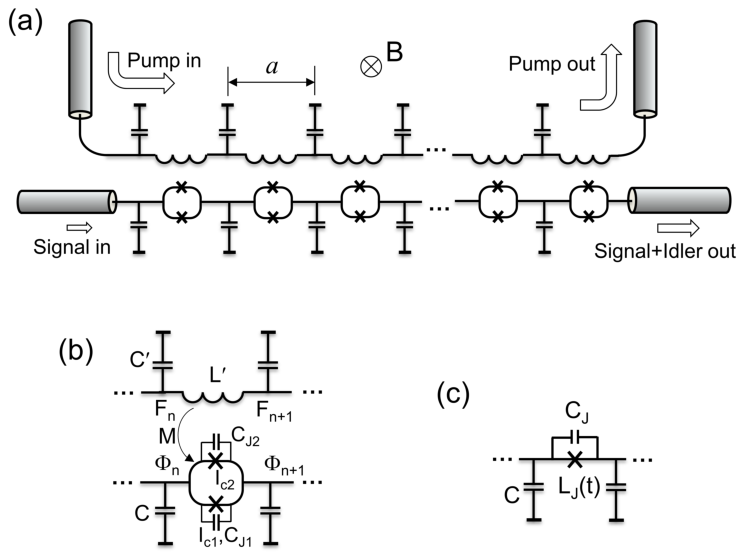}
\caption{(a) Architecture of a TWJPA driven by means of a flux wave traveling in a separate line.
The pump and signal transmission lines consist of similar number $N$ $(\gg 1)$ of identical elementary
cells having similar length $a$. A homogeneous static magnetic field $\textbf{B}$ applied perpendicular
to the circuit plane sets the constant flux bias $\Phi_{\textrm{dc}}$ and allows adjustment of the SQUID inductances.
(Alternatively, the flux bias can be set by a constant feeding current in the pump transmission line.)
(b) The pair of inductively coupled elementary cells in the pump and signal lines.
(c) Equivalent circuit of the signal-line cell,
including ground capacitance $C$ and an effective Josephson junction with self-capacitance $C_J$
and time-dependent linear Josephson inductance $L_J(t)$.} \label{EqvSchm}
\end{center}
\end{figure}

\begin{figure}[t]
\begin{center}
\includegraphics[width=3.0in]{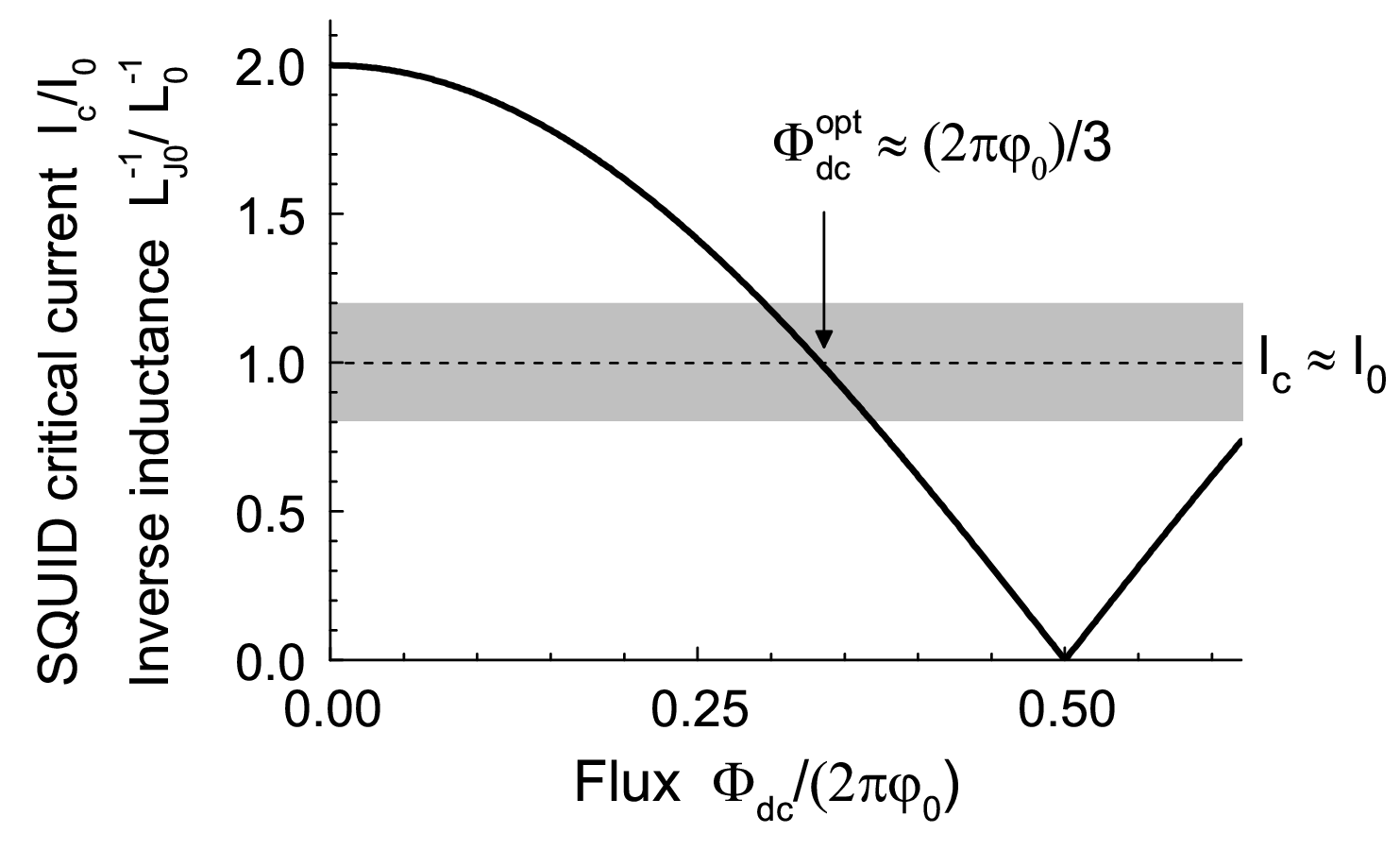}
\caption{Inverse inductance $L^{-1}_{J0}$ (or critical current $I_c$) as a function
of the constant flux bias. The shaded band corresponds to the $\pm$20\% deviation from
the optimal value $1/L_{0} \approx 1/L'$ allowing both sufficiently large flux modulation
of the inductance and a sufficiently wide range for fine adjustment of
the phase velocities in two lines (see Eq.~(\ref{cutoff-equal})).}
\label{inverseInd}
\end{center}
\end{figure}

\section{Wave equation and flux drive}

The electric diagram of our superconducting microwave circuit with nominally
vanishing losses is shown in Fig.~1.
The circuit consists of two inductively coupled ladder-type transmission lines:
a signal line and a pump line.
Similarly to the design of the cavity-based JPA studied in Ref. \cite{Castellanos-Beltran2008},
the cells of our signal line include symmetric dc SQUIDs ($I_{c1} = I_{c2} = I_0$) with
small geometrical inductances of the loops. In this case, the effective critical current
$I_c = 2I_{0}|\cos(\Phi/2\varphi_0)|$ and the corresponding linear inductance of
each SQUID,
\begin{equation}
L^{-1}_{J0} = (2 I_0/\varphi_0)  |\cos(\Phi/2\varphi_0)|,
\label{dcSQUID}
\end{equation}
can be efficiently controlled by magnetic flux
$\Phi = \Phi_{\textrm{dc}} + \Phi_{\textrm{ac}}$.
These SQUIDs are interleaved with
identical ground capacitances $C$ (Fig.~1b).
The pump transmission line consists of $LC$ sections with inductances $L'$ and
ground capacitances $C' \approx C$; its impedance
$Z_0 = \sqrt{L'/C'} \approx 50~\Omega$. The coupling inductances between
SQUIDs and inductors $L'$ are equal to $M~ (\ll L')$ and enable
ac flux drive in the SQUIDs.

The flux offset $\Phi_{\textrm{dc}}$ is set by means of an external static field $\textbf{B}$
and ensures the SQUID critical current $I_c \sim I_{0}$ and hence the SQUID
inductance $L_{J0}(\Phi_{\textrm{dc}}) \sim L_0 \equiv \varphi_0/I_0$.
Then, as shown in Fig.~2, the ac component of the flux $\Phi_{\textrm{ac}}$ can result in
sufficiently large modulation of the inductance, i.e., $0 < L^{-1}_{J0} < 2L^{-1}_0$.
Therefore, efficient three-wave mixing is possible
in such a SQUID without use of the nonlinear properties of its inductance.
``Fine-tuning" of the flux $\Phi_{\textrm{dc}}$ and therefore the SQUID inductance $L_{J0} \approx L'$
allows adjustment of frequencies $\omega_0$ and $\omega'_0$,
\begin{equation}
\omega_0 = 1/\sqrt{L_{J0}C} \approx \omega'_0 = 1/\sqrt{L'C'},
\label{cutoff-equal}
\end{equation}
keeping the signal-line impedance $\sqrt{L_{J0}/C} \approx \sqrt{L'/C'} \approx 50~\Omega$.
Ignoring for the moment small dispersion, Eq.~(\ref{cutoff-equal}) ensures that the phase
velocities of the waves in the signal line,
\begin{equation}
v_s = a \omega_0,
\label{v-s}
\end{equation}
and in the pump line,
\begin{equation}
v_p = a \omega'_0,
\label{v-p}
\end{equation}
are equal, where $a$ is the size of the cells in each transmission line.
Therefore, spatial phase matching is possible, i.e., $\Delta k =k_p - k_s - k_i \approx 0$.

Generally, the dynamics of the electrical circuit shown in Fig.~1a is described by the set of
coupled equations for the values of fluxes $F_n$ on the nodes of the pump line
and fluxes $\Phi_n$ on the nodes of the signal line (see Fig.~1b). These equations
are derived in Appendix A.
In the case of sufficiently small dimensionless coupling, i.e.,
\begin{equation}
\kappa = M/L' \ll 1,
\label{kappa}
\end{equation}
and sufficiently large power $P_p$, the pump
remains undepleted and presents a traveling flux wave having constant
amplitude $A_{p0} \propto \sqrt{P_p}$.
Then, the equations for these two transmission lines decouple. Eventually, the equation
of motion for the signal line in the continuum limit (valid for $\omega \ll \omega_0, \omega_J$)
reads
\begin{eqnarray}
\frac{\partial^2 \phi}{\partial x^2} - \omega^{-2}_0\frac{\partial^2 \phi}{\partial t^2}
+ \omega^{-2}_J \frac{\partial^4 \phi}{\partial t^2\partial x^2}
-  \gamma \frac{\partial}{\partial x}
\left[ \left( \frac{\partial \phi}{\partial x} \right)^3 \right] \nonumber \\
+ \frac{\partial}{\partial x} \left[ m \sin(k_p x - \omega_p t) \frac{\partial \phi}{\partial x} \right]  = 0.~~
\label{Eq-motion-signal}
\end{eqnarray}
Here, $x$ is the dimensionless coordinate normalized on the cell size $a$; the normalized
magnetic flux $\phi(x,t) = \Phi(x,t)/\varphi_0$, where the continuous
flux variable $\Phi(x,t)$ coincides with flux values $\Phi_n$ on the nodes
$x = n$. Frequency
\begin{equation}
\omega_J = 1/\sqrt{L_{J0}C_J}
\label{w-plasma}
\end{equation}
is the plasma frequency of the SQUID, which
has total capacitance $C_J = C_{J1} + C_{J2}$ (see Fig.~1b) and effective inductance $L_{J0}$.
The dimensionless wave numbers of the pump wave in the pump line, $k_p$, and of any wave
(including signal $\omega_s$ and idler $\omega_i$) in the signal line, $k$,
obey the dispersion relations (see Appendix A)
\begin{eqnarray}
k_{p} \approx \frac{\omega_p}{\omega'_0} \left( 1 + \frac{\omega_p^2}{24\omega'^2_0} \right)
\approx \frac{\omega_p}{\omega'_0}  \ll 1
\label{disp-relation-approx-p}
\end{eqnarray}
and
\begin{eqnarray}
k &\approx& \frac{\omega}{\omega_0} \left( 1 + \frac{\omega^2}{2\omega^2_{J}}
+ \frac{\omega^2}{24\omega^2_0} \right) \nonumber \\
&\approx& \frac{\omega}{\omega_0} \left( 1 + \frac{\omega^2}{2\omega^2_{J}}\right) \ll 1,
\label{disp-relation-approx}
\end{eqnarray}
respectively. The latter relation includes the small
term $\omega^2/2\omega^2_J$
which stems from the Josephson plasma resonance in the SQUIDs. \cite{Yaakobi2013}
The small terms $\omega_{p}^2/24\omega'^2_0$ and $\omega^2/24\omega^2_0$
in Eqs.~(\ref{disp-relation-approx-p}) and (\ref{disp-relation-approx}), respectively,
are merely the consequence of the fact that the transmission lines are made
of lumped elements. In the practical case
of $\omega^2_J \lesssim \omega^2_0$, these small
terms can be omitted.

The fourth, Kerr-like term (with coefficient $\gamma = 1/6$) on the left-hand
side of Eq.~(\ref{Eq-motion-signal}) describes
nonlinear effects, \cite{Yaakobi2013} which are normally negligibly
small (unless the signal amplitude or, more specifically,
the Josephson phase difference on the SQUIDs approaches
appreciable values, $\varphi(x) = k_{s,i} \Phi(x)/\varphi_0 \sim 1$; see Section V for more details).
For reasonably small input signal, the signal wave can approach such large amplitude only
at the transmission-line output.  So counting of this term is important only for evaluation
of the gain compression.

The fifth term on the left-hand side of the wave equation (\ref{Eq-motion-signal})
stems from the time-dependent distributed inductance $L(x,t)$, which is
varied in a traveling-wave fashion:
\begin{equation}
L^{-1}(x,t) =[1 + m \sin(k_p x - \omega_p t)] L^{-1}_{J0}.
\label{SQUID-distr-inductance}
\end{equation}
The dimensionless coefficient $m$,
\begin{equation}
m = \frac{\kappa}{2}A_{p0} k_p \tan \frac{\Phi_{\textrm{dc}}}{2 \varphi_0}
\approx \kappa \frac{\sqrt{Z_0 P_p}}{\sqrt{2}\varphi_0 \omega_0} \tan \frac{\Phi_{\textrm{dc}}}{2 \varphi_0},
\label{m-value-def}
\end{equation}
determines the depth of the inductance modulation
and depends on the strength of coupling $\kappa$
and the pump power $P_p$.
Here the product $A_{p0} k_p $ is equal to
$A_{p0}\omega_p/\omega_0 = V_p/\varphi_0\omega_0$, where $V_p = \sqrt{2Z_oP_p}$ is the
voltage amplitude in the pump wave.

The time-dependent inductance (\ref{SQUID-distr-inductance}) enables parametric
amplification of the signal, \cite{Tien1958, Cullen1960, Boyd2008} given in the
ideal case by  \cite{White2015,Zorin2016}
\begin{eqnarray}
G &=&\frac{P^\textrm{out}_s}{P^\textrm{in}_s} =\cosh^2 g N \approx \frac{1}{4} e^{2 g N},  \label{gain-exp} \\
G_i &=& \frac{P^\textrm{out}_i}{P^\textrm{in}_s} =\frac{\omega_i}{\omega_s}\sinh^2 g N \approx \frac{\omega_i}{4\omega_s} e^{2 g N},
\label{interm-gain-exp}
\end{eqnarray}
where both the direct power gain $G~ (\gg 1)$ and the intermodulation
power gain $G_i~ (\gg 1)$
depend exponentially on length $N$ and the gain factor $g \propto m \omega_p/\omega_0$.
Its inverse value,
\begin{equation}
g^{-1} \equiv N_g,
\label{e-folding-length}
\end{equation}
is therefore the $e$ folding length of the amplitude gain.

\section{Phase-matching consideration}

In the general case, an inherently broadband flux-driven transmission line
allows multiwave-mixing processes.
Specifically, the basic three-wave mixing process involving
conventional idler frequency
\begin{equation}
\omega_i = \omega_p - \omega_s,
\label{3WM conventional}
\end{equation}
can be accompanied by two additional mixing processes (frequency up-conversion)
involving the idlers~  \cite{Erickson2017}
\begin{eqnarray}
\omega_{1} &=& \omega_p + \omega_s,  \label{3WM additional0}
\end{eqnarray}
and
\begin{eqnarray}
\omega_{2} &=& \omega_p + \omega_i = 2\omega_p - \omega_s.
\label{3WM additional 2}
\end{eqnarray}
The latter expression for idler frequency $\omega_2$ allows us to interpret
the process in Eq.~(\ref{3WM additional 2}) as a four-wave mixing.
These three types of mixing [Eqs.~(\ref{3WM conventional}),
(\ref{3WM additional0}), and (\ref{3WM additional 2})],
are schematically shown in Fig.~3a.

\begin{figure}[t]
\begin{center}
\includegraphics[width=3.3in]{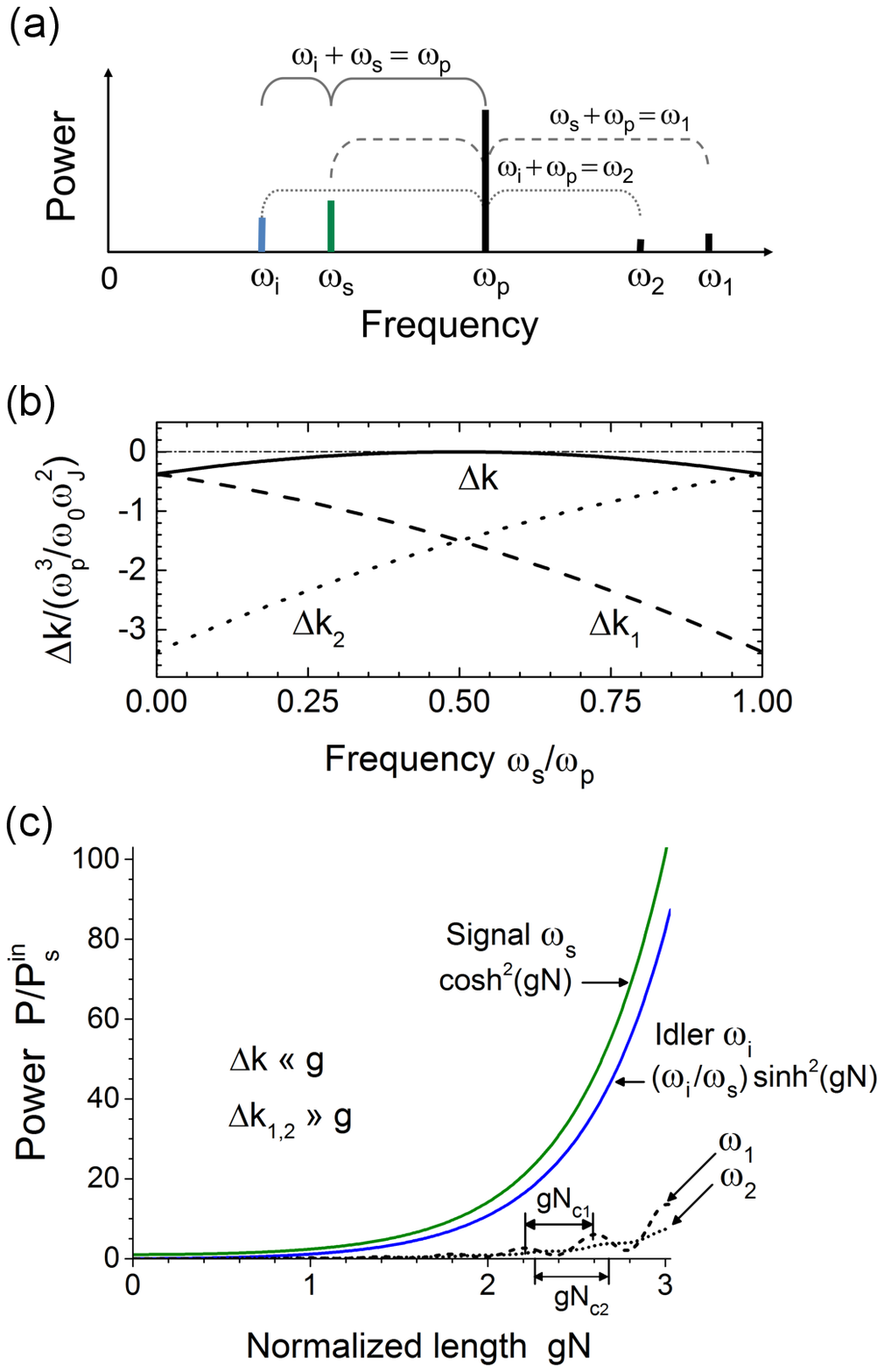}
\caption{(a) The mixing processes
involving the pump frequency ($\omega_p$), signal frequency ($\omega_s$) and one of three idler
frequencies; that is $\omega_i$ (solid brackets), $\omega_1$ (dashed brackets),
and $\omega_2$ (dotted brackets).
(b) Normalized phase mismatches for the mixing processes (\ref{3WM conventional}),
(\ref{3WM additional0}), and (\ref{3WM additional 2}) in the case of optimal
dc flux offset (ensuring fulfilment of the relation $\omega_0 - \omega'_0 = \omega^2_p \omega_0/8\omega^2_J$)
as a function of the reduced signal frequency $\omega_s/\omega_p$.
(c) Normalized power of the signal $\omega_s$ given by Eq. (\ref{gain-exp}) (solid green line),
the idler $\omega_i$ given by Eq. (\ref{interm-gain-exp}) (solid blue line)
and the idlers $\omega_1$ (dashed line) and $\omega_2$ (dotted line),
as a function of the TWJPA length.
Negligibly small phase mismatch $\Delta k$ for the basic process
(see Eq.~(\ref{cohlength-0}))
and large phase mismatches $\Delta k_{1} $ and $\Delta k_{2} $ 
for the unwanted mixing processes
(see Eq.~(\ref{cohlength-12})) are assumed. For clarity, the signal frequency is
close to $\omega_p/2$ (i.e., $\omega_s=0.55 \omega_p$.)}.
\label{mismatch-and-gain}
\end{center}
\end{figure}

If the processes in expressions (\ref{3WM additional0}) 
and (\ref{3WM additional 2}) 
are not sufficiently suppressed, 
they may degrade the amplifier performance.
In this general case, the Manley-Rowe relations \cite{Manley-Rowe1956}
for a pure three-wave-mixing process [Eq.~(\ref{3WM conventional})]
are modified.
Specifically, the original relations for the wave powers, \cite{Boyd2008}
\begin{eqnarray}
\frac{P_s - P^\textrm{in}_s}{\omega_s} = \frac{P_i}{\omega_i} = - \frac{P_p -P^\textrm{in}_p}{\omega_p},
\label{3WM-MR-0}
\end{eqnarray}
or, equivalently, the relations for the corresponding photon numbers,
\begin{equation}
n_s - n_s^\textrm{in} =  n_i =n_p^\textrm{in} - n_p,
\label{3WM-MR-0-n}
\end{equation}
take the form
\begin{eqnarray}
\frac{P_s - P^\textrm{in}_s}{\omega_s}
&=& \frac{P_i}{\omega_i}-\frac{P_1}{\omega_1} +\frac{P_2}{\omega_2} \nonumber \\
&=& -\frac{P_p -P^\textrm{in}_p}{\omega_p} - \frac{2P_1}{\omega_1}- \frac{P_2}{\omega_2},
\label{3WM-MR-02}
\end{eqnarray}
or, in terms of the photon numbers,
\begin{equation}
n_s  - n_s^\textrm{in} = n_i - n_1 + n_2= n_p^\textrm{in}-n_p-  2n_1 - n_2,
\label{3WM-MR-02-n}
\end{equation}
where $P_{1,2}$ and $n_{1,2}$ are the powers and the photon numbers
of the modes $\omega_{1,2}$, respectively.
Because of large photon energies, $\hbar\omega_{1,2} >\hbar \omega_p$,
both complementary processes, $\omega_{s,i} + \omega_p \rightarrow \omega_{1,2}$
and $\omega_{1,2} \rightarrow \omega_{s,i} + \omega_p$,
are possible, \cite{Yariv1967}
so the amplified power of the signal/idler flows repeatedly from
mode $\omega_{s,i}$ to mode $\omega_{1,2}$ and vice versa.~\cite{Boyd2008,Yariv1967}
This scenario leads to reduction of the direct (intermodulation)
gain $G=P^\textrm{out}_{s}/P^\textrm{in}_s$ ($G_i=P^\textrm{out}_{i}/P^\textrm{in}_s$).
As we show below,
suppression of unwanted modes $\omega_1$ and $\omega_2$ is possible
by properly choosing the circuit parameters, i.e., without modification
of the circuit architecture.

Using expression (\ref{disp-relation-approx-p}) for pump-wave number $k_p$ and
expression (\ref{disp-relation-approx})
for wave numbers $k_s$, $k_i$, $k_1$, and $k_2$ of the waves $\omega_s$, $\omega_i$, $\omega_1$,
and $\omega_2$, respectively,
the corresponding phase mismatches for the
processes (\ref{3WM conventional}), (\ref{3WM additional0}), and (\ref{3WM additional 2})
can be written as
\begin{eqnarray}
\Delta k &=& k_p - k_i - k_s = \nu - (1/3+ \delta^2)\eta, \label{dk0} \\
\Delta k_{1} &=& k_p - k_1 + k_s  = \nu - (13/3 + 4 \delta + \delta^2)\eta, \label{dk1} \\
\Delta k_{2} &=& k_p - k_2 + k_i = \nu - (13/3 - 4 \delta + \delta^2)\eta.
\label{dk2}
\end{eqnarray}
Here the dimensionless frequency detuning is
\begin{equation}
\delta = \frac{2\omega_s - \omega_p}{\omega_p} = \frac{\omega_p - 2\omega_i }{\omega_p}.
\label{delta-def}
\end{equation}
In Eqs. (\ref{dk0}), (\ref{dk1}), and (\ref{dk2}),
we have introduced two small parameters:
\begin{equation}
\eta = 3\omega^3_p/8\omega_0\omega^2_J
\label{eta-def}
\end{equation}
and
\begin{equation}
\nu = \omega_p/\omega'_0 - \omega_p/\omega_0 \approx \Delta\omega \:\omega_p/\omega^2_0 ,
\label{nu-def}
\end{equation}
where $\Delta\omega = \omega_0 - \omega'_0 \ll \omega_0$
is a small mismatch of the cutoff frequencies in two transmission lines
(see Eq.~(\ref{cutoff-equal})).
Therefore, properly designing the plasma frequency of the
SQUIDs $\omega_J$ (see Eq.~(\ref{w-plasma})) and setting a small cutoff-frequency difference,
$\Delta\omega > 0$,
allow effective control of all three phase mismatches (\ref{dk0}), (\ref{dk1}) and (\ref{dk2}).
Specifically, fixing the frequency difference as
\begin{equation}
\Delta\omega = \omega^2_p \omega_0/8\omega^2_J,
\label{delta-omega}
\end{equation}
or, equivalently, taking
$\nu = \eta/3$, one can achieve nearly perfect phase matching ($\Delta k \approx 0$)
in the basic mixing process (\ref{3WM conventional}) for frequencies
$\omega_s \approx \omega_i \approx \omega_p/2$ (or small
detuning $|\delta| \ll 1$), while keeping appreciable
phase mismatches
in unwanted processes (\ref{3WM additional0})
and (\ref{3WM additional 2}): that is,
\begin{eqnarray}
\Delta k &=& - \delta^2\eta, \label{dk0mod} \\
\Delta k_{1} &=& -(4 + 4 \delta + \delta^2)\eta, \label{dk1mod} \\
\Delta k_{2} &=&  -(4 - 4 \delta + \delta^2)\eta.
\label{dk2mod}
\end{eqnarray}
The dependencies of these values on the normalized signal frequency
are shown in Fig.~3b.

In the case of sufficiently small $|\Delta k |$, or, equivalently,
large coherence length $N_c$ in the basic process,
\begin{equation}
N_c \equiv \frac{\pi}{|\Delta k|} = \frac{\pi}{\delta^2\eta} \gg \frac{1}{g} = N_g,
\label{cohlength-0}
\end{equation}
and sufficiently short coherence lengths $N_{c1}$ and $N_{c2}$ in two additional mixing
processes, i.e.,
\begin{equation}
N_{c1,c2} \equiv \frac{\pi}{|\Delta k_{1,2}|} \approx \frac{\pi}{4\eta} \ll \frac{1}{g} = N_g,
\label{cohlength-12}
\end{equation}
the unwanted processes (\ref{3WM additional0}) and (\ref{3WM additional 2}) are safely suppressed.
So amplification of the signal ($\omega_s$) and
idler ($\omega_i$) frequencies is described by
Eqs.~(\ref{gain-exp}) and (\ref{interm-gain-exp}),
respectively.  This favorable case is illustrated in Fig.~3c.

\section{Signal gain}

To quantify the effect of multiwave mixing (shown schematically in Fig.~3a) we apply the coupled-mode
equation method \cite{Agrawal}. The solution of Eq.~(\ref{Eq-motion-signal}) is then found
in the form of four waves propagating in the forward direction,
\begin{eqnarray}
\phi(x,t) = \frac{A_s(x)}{2}  e^{i(k_sx-\omega_s t)}+ \frac{A_i(x)}{2} e^{i(k_ix-\omega_i t)} \nonumber \\
+ \frac{A_1(x)}{2}  e^{i(k_1x-\omega_1 t)}+\frac{A_2(x)}{2}  e^{i(k_2x-\omega_2 t)}+\textrm{c.c.},
\label{solution}
\end{eqnarray}
where $A_{s}(x)$, $A_{i}(x)$, $A_{1}(x)$, and $A_{2}(x)$ are slowly varying complex amplitudes of the signal
and three idlers, obeying the condition \cite{Yaakobi2013,Zorin2016}
\begin{equation}
\left| \frac{\partial^2A_{s,i,1,2}}{\partial x^2} \right| \ll k_{s,i,1,2} \left| \frac{\partial A_{s,i,1,2}}{\partial x} \right|
\ll k_{s,i,1,2}^2 \left| A_{s,i,1,2} \right|.
\label{assumption}
\end{equation}
After incorporation of Eq.~(\ref{solution}) into Eq.~(\ref{Eq-motion-signal}) with the omitted Kerr term ($\propto \gamma$),
the coupled linear equations for $A_{s}$, $A_{i}$, $A_1$, and $A_2$, take the form
\begin{eqnarray}
\frac{dA_s}{dx} &=&  \frac{m}{2} k_i A^*_i e^{i\Delta k x}  + \frac{m}{2} k_1 A_1 e^{-i\Delta k_1 x},  \label{cme-s} \\
\frac{dA_i}{dx} &=&  \frac{m}{2} k_s A^*_s e^{i\Delta k x} + \frac{m}{2} k_2 A_2 e^{-i\Delta k_2 x}, \label{cme-i} \\
\frac{dA_1}{dx} &=&  - \frac{m}{2} k_s A_s e^{i\Delta k_1 x}, \label{cme-1} \\
\frac{dA_2}{dx} &=& - \frac{m}{2} k_i A_i e^{i\Delta k_2 x}. \label{cme-2}
\end{eqnarray}
The first terms on the left-hand-sides of Eqs.~(\ref{cme-s})
and (\ref{cme-i}), 
describe the basic parametric mixing (\ref{3WM conventional}). \cite{Agrawal}

In the case of sufficient suppression of modes $\omega_1$ and $\omega_2$
(see Eq.~(\ref{cohlength-12}) or, equivalently, the inequality $\pi g/4\eta \ll 1$)
small amplitudes $A_{1}$ and $A_{2}$ can be omitted and the decoupled pair of equations (\ref{cme-s})
and (\ref{cme-i})  for solely amplitudes $A_{s,i}$ has a simple analytical solution.
Specifically, the solution with initial conditions  $A_s(0) = A_{s0}$ and
$A_i(0) = 0$ reads \cite{White2015,Zorin2016}
\begin{eqnarray}
A_s &=& A_{s0} \left[\cosh g x - i \frac{\Delta k}{2g}
\sinh g x\right] e^{i \Delta k x/2} ,~~ \label{signal-solution} \\
A_i &=& 2 \frac{g_0 \omega_s}{g \omega_p} A_{s0} \sinh (g x) e^{i \Delta k x/2}. \label{idler-solution}
\end{eqnarray}
Here the exponential gain factor is
\begin{equation}
g = \sqrt{(1-\delta^2) g_0^2- (\Delta k/2)^2 },
\label{gain}
\end{equation}
and its maximum value,
\begin{equation}
g_0  = \frac{m k_p}{4} \approx  \frac{m \omega_p}{4\omega_0},
\label{max-gain}
\end{equation}
is achieved at perfect phase matching, $\Delta k =0$, and zero detuning, $\delta =0$.

\begin{figure}[b]
\begin{center}
\includegraphics[width=3.5in]{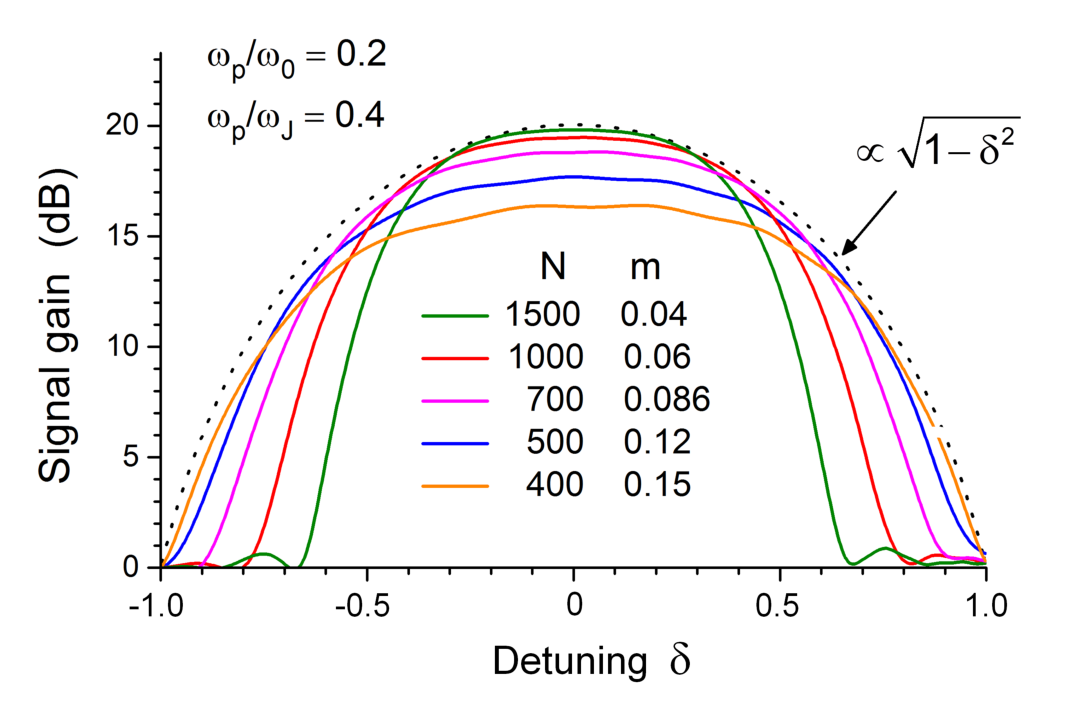}
\caption{Signal gain as a function of frequency $\omega_s$ at fixed pump
frequency $\omega_p = 0.2 \omega_0$ and Josephson plasma frequency $\omega_J = 0.5 \omega_0$
(yielding $\eta = 0.012$, see Eq.~(\ref{eta-def})). Solid curves correspond to
several combinations of length $N$ (in decreasing order from top to bottom) and
modulation parameter $m$ with fixed
product $g_0N = m\omega_p N/4\omega_0 = 3$ (yielding nominal maximum gain of $G = \cosh^2 g_0N \approx 10^2$).
The dotted curve shows the ultimate gain versus frequency
given by Eq.~(\ref{max-power-gain}).}
\label{mismatch}
\end{center}
\end{figure}

The power gain $G$ of the line of length $N$ is then given by formula~\cite{White2015}
\begin{equation}
G=\frac{|A_s(N)|^2}{|A_{s0} |^2} = \cosh^2 g N - \left(\frac{\Delta k}{2g}\right)^2 \sinh^2 g N,
\label{general-case-power-gain}
\end{equation}
which for zero phase mismatch, $\Delta k =0$, and large gain, $g N \gg 1$,
is reduced to Eq.~(\ref{gain-exp})
with $g = g_0 \sqrt{1-\delta^2}$, i.e.,
\begin{equation}
G = \cosh^2(g_0 \sqrt{1-\delta^2} N) \approx \frac{1}{4} \exp \left( \frac{\omega_p \sqrt{1-\delta^2}}{2\omega_0} m N  \right).
\label{max-power-gain}
\end{equation}
The frequency dependence of $\log G$
has a typical semicircle shape, $y =\sqrt{1-\delta^2}$, with the maximum
at $\delta =0$. \cite{Tien1958,Cullen1960}
For example, the amplifier with this ideal shape of the power gain with maximum value
of 20~dB has a 3-dB bandwidth of
$\sqrt{1-(17/20)^2}\, \omega_p \approx  0.5 \omega_p$, which is centered at $\omega_s = 0.5\omega_p$
(see the dotted line in Fig.~4).~\cite{Cullen1960,Zorin2016}.

In the case of insufficiently short coherence lengths  $N_{c1}$ and $N_{c2}$ [i.e. only
slightly less than or approximately equal to $N_g$;
see Eq.~(\ref{cohlength-12})], that is, 
a not sufficiently small value of the dimensionless parameter
\begin{equation}
\xi = \frac{N_{c1,c2}}{N_g}
= \frac{\pi g_0 \sqrt{1-\delta^2}}{4 \eta} \approx \frac{m \omega^2_J \sqrt{1-\delta^2}}{2 \omega^2_p}  \lesssim 1,
\label{small-parameter}
\end{equation}
the power conversion to the modes $\omega_1$ [Eq.~(\ref{3WM additional0})]
and $\omega_2$ [Eq.~(\ref{3WM additional 2})] becomes appreciable.
So, redistribution of power between the signal, the idler and the unwanted idler modes $\omega_1$ and $\omega_2$
may lead to substantial suppression of the signal gain.
To quantify this effect, which is most pronounced at large detuning, $|\delta| \geq 0.5$,
we numerically solve the set of differential
equations (\ref{cme-s})-(\ref{cme-2}).

Figure 4 shows the frequency dependence of $\log G$ obtained by
numerical integration of Eqs.~(\ref{cme-s})-(\ref{cme-2})
with initial conditions
\begin{equation}
A_s(0) = A_{s0},~~~ A_i(0) = A_1(0) =  A_2(0) = 0,
\label{initial-conditions}
\end{equation}
for five TWJPAs having different lengths $N$ (the set of solid curves),
but nominally the same maximum gain of 20~dB (adjusted by a proper
pump power, $P_p \propto 1/N^2$, keeping $m \propto 1/N$).
We find $\xi$ is approximately equal
to 0.13, 0.19, 0.27, 0.38, and 0.47 (in the order of the curves from top to bottom
in the central part of the plot)
which cover the range from the relatively small value of 0.13 up
to the value 0.47 close to unity.
For comparison, the dotted curve illustrates the case of $\xi = 0$,
yielding the maximum possible gain given by Eq.~(\ref{max-power-gain}).

One can see two features in the behavior of these circuits.
Firstly, the gain suppression in the TWJPAs with relatively small values
of $\xi$  ($\xi < 0.3$ or $m < 0.1$) is rather small ($< 2$ dB) for
detuning $|\delta| \lesssim 0.5$, whereas for
$0.5 \lesssim |\delta| \leq 1$ the gain suppression is significant. This behavior is due to increase
of the phase mismatch $|\Delta k| = \delta^2 \eta$ and hence violation
of inequality (\ref{cohlength-0}) occurring at relatively large $|\delta|$.
Secondly, the signal gain for rather large values of $\xi$ ($\xi > 0.3$ or $m > 0.1$)
is notably less than the nominal value given by the dotted line,
especially for small detuning, $|\delta| \lesssim 0.5$.
That suppression of the gain for larger detuning, $|\delta| \gtrsim 0.5$, is
not so dramatic as in the case of small $\xi$.
So, we can conclude that for our set of characteristic frequencies,
a TWJPA length $N$ of about 1000 ensures both
sufficiently large bandwidth of about $0.47 \omega_p$ and sufficiently
small unwanted power conversion to the high-frequency idler modes.

\section{Possible circuit design}

The design of the practical circuit obviously depends on
specific applications of the TWJPA.  Below we focus on the set of parameters
typical for the TWJPAs (both with three-wave mixing and four-wave mixing)
that were earlier developed for the quantum-information
applications including superconducting qubits.  \cite{OBrien2014,White2015,Macklin2015,Bell-Samolov2015,Zorin2016,Zorin2017,Miano2018,WenyuanZhang2017}
These parameters include an operation frequency on the order of 10 GHz,
relatively large Josephson junctions having critical current on the order
of several microamperes (and hence sufficiently low charging energy $E_c = e^2/2C_J
\sim 10^{-3}\varphi_0 I_0$, ensuring classical behavior  of the Josephson phase $\varphi$),
a standard line impedance of 50~$\Omega$, and a nominal gain of 20~dB.

Taking a relatively high target value of the cutoff frequency $\omega_0/2\pi = 100$~GHz
and the standard transmission-line impedance $Z_0 = 50~\Omega$,
we obtain the following basic circuit parameters,
\begin{equation}
C' = C = 1/\omega_0 Z_0 \approx 32~ \textrm{fF}
\label{C-circuit-parameters}
\end{equation}
and
\begin{equation}
L' = L_{J0} = Z_0/\omega_0 \approx 80~ \textrm{pH}.
\label{L-circuit-parameters}
\end{equation}
For a typical size of an elementary cell $a \sim 30~\mu$m (see possible circuit layout,
for example, in Ref.~\cite{Zorin2017}) the phase velocity of microwaves in these
transmission lines $v_s = a \omega_0$ is reduced
to about 2$\times 10^7$ m/s.
The wavelength of the pump $\lambda_p$ for the designed frequency of $\omega_p/2\pi=20$~GHz
(i.e., $\omega_p/\omega_0 = 0.2$)
is equal to the length of $N_p = 2\pi \omega_0/\omega_p
\approx 31$ elementary cells.

In the optimal working point the constant magnetic flux
$\Phi^{\textrm{opt}}_{\textrm{dc}}/2\varphi_0 \approx \pi/3$
yields about 50~\% suppression of the maximum critical current of the SQUIDs,
so the Josephson inductance [Eq.~(\ref{L-circuit-parameters})]
corresponds to the critical current of a single junction, i.e.,
$I_0 = \varphi_0/L_{J0} \approx 4~\mu$A (see Fig.~2).
Such Josephson junctions can be fabricated using either multilayer niobium technology (see, e.g.,
Refs.~\cite{Dolata2005,Tolpygo2017}) or shadow-evaporation aluminum technology. \cite{Niemeyer1974}
The Josephson plasma frequency of a bare Josephson junction with critical current
density of, say, $j_c = 3~ \mu \textrm{A}/\mu \textrm{m}^2$ and specific barrier capacitance
about $c_J=50$~fF/$\mu\textrm{m}^2$ is $ \sqrt{j_c/\varphi_0 c_J}/2\pi \approx 70$~GHz.
\cite{van-der-Zant1994} Therefore, the plasma
frequency of the flux-biased SQUID with a partially suppressed critical
current is $\omega_J/2\pi \approx 70\sqrt{\cos(\pi/3)}~\textrm{GHz} \approx 50$~GHz,
i.e., $\omega_p/\omega_J = 0.4$.
Taking the total number of elementary cells $N= 1000$, one can achieve sufficiently large
gain in a reasonably large bandwidth (see Fig.~4).
Possible deterioration of performance due to imperfect flux setting and inhomogeneity
of the transmission-line parameters (including those due to regular and irregular
inhomogeneities of the critical current density on the chip)
should be sufficiently small. The corresponding estimations are given in Appendix B.

According to formula (\ref{m-value-def}),
the target value of the modulation parameter $m = 0.06$
can be achieved by engineering a reasonably small dimensionless coupling $\kappa = M/L'$ (e.g.,
on the order of 0.02) and applying sufficiently large pump power,
\begin{equation}
P_p = 2  \cot^2(\pi/3)  \frac{(m \varphi_0 \omega_0)^2}{\kappa^2 Z_0} \approx -53~\textrm{dBm}.
\label{pump-power}
\end{equation}
This power passing though the superconducting transmission line is not
immediately dissipated on the chip, but can cause some heating of cold attenuators
installed in the base-temperature stage.
However, this level of power is seemingly acceptable for the most of setups with dilution
refrigerators.

In conventional TWJPAs operating on the basis of wave mixing with the aid of
Josephson nonlinearity, a relatively small pump power (limited by the nonlinear-element characteristics,
i.e., the critical current) propagates together with the signal and idler waves in the common
transmission line. Because of parametric interaction of these traveling waves the pump power
is gradually converted into the signal and the idler. The resulting pump depletion leads
to gradual reduction of the signal gain and, finally, to gain saturation. This effect
dramatically limits the amplifier dynamic range. \cite{OBrien2014,Zorin2016}
The important feature of the proposed TWJPA driven
by relatively large power (\ref{pump-power}) fed into a separate port of an isolated
$LC$ line and only partly converted into the signal and idler is almost constant pump power
(i.e., the absence of pump depletion). This remarkable property enables
ultimate mixing in all section of the signal line and, hence, a steady gain.
Only at sufficiently large signal amplitude achieved in the output sections of the line
(i.e., for ac amplitude amounting to about $I_0$) is its further growth impeded
by the SQUID nonlinearity.

\begin{figure}[b]
\begin{center}
\includegraphics[width=3.5in]{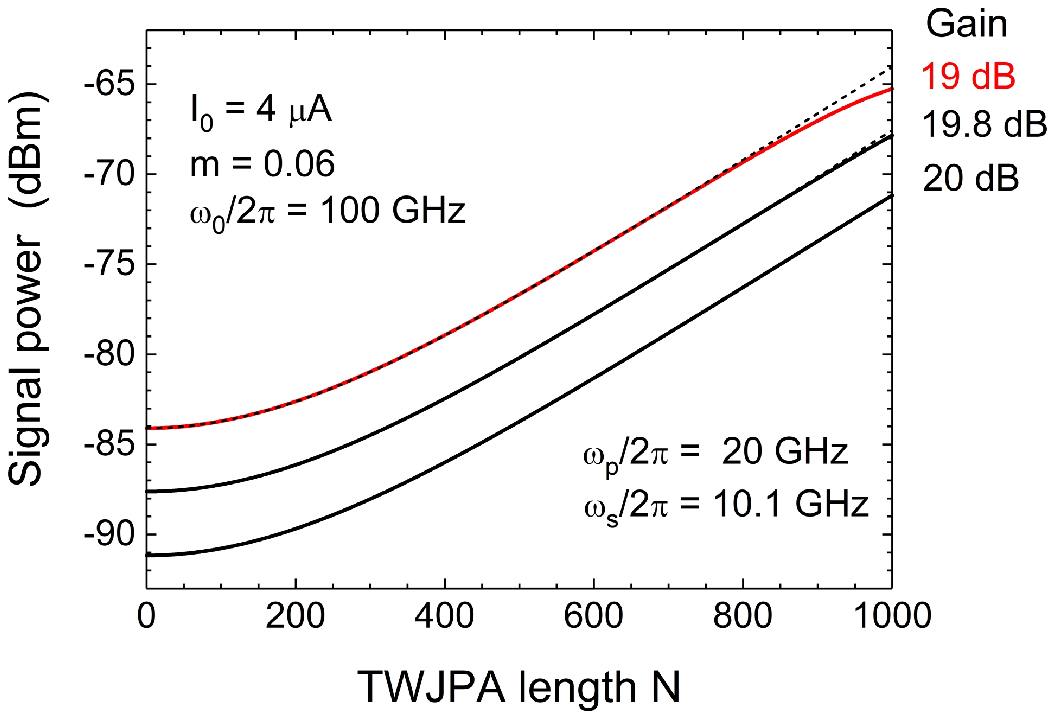}
\caption{Increasing power of the signal waves of frequency $\omega_s = 0.505 \omega_p$
with input powers of $-91$ dBm, $-87.7$ dBm,
and $-84$ dBm (solid lines in the order from bottom to top, respectively) as
a function of the cell number [found by numerical solution of the coupled-mode
nonlinear equations (\ref{cme-s-gamma}) and (\ref{cme-i-gamma})].
The actual gain for different input powers
is given in the right column.
The topmost solid curve (red) corresponds to decrease of the nominal 20-dB gain by 1 dB,
which occurs at input power $P_{1\textrm{dB}}=-84$ dBm.
For comparison, the dashed lines show the signal-power
growth ignoring the nonlinear effects
($\gamma = 0$); that is, applying the formula $P_s = P^\textrm{in}_s \cosh^2 g_0 N$
(see Eq.~(\ref{max-power-gain})). }
\label{compression}
\end{center}
\end{figure}

To evaluate this effect in the limit of pure three-wave mixing ($|A_{1,2}| \ll |A_{s,i}|$),
we derive from Eq.~(\ref{Eq-motion-signal}) the pair of coupled nonlinear equations
for complex amplitudes $A_s$ and $A_i$:
\begin{eqnarray}
\frac{dA_s}{dx} =  \frac{m}{2} k_i A^*_i e^{i\Delta k x}
+ i\frac{3}{8} \gamma k_s A_s ( k_s^2 |A_s|^2 + 2 k_i^2 |A_i|^2),\;  \label{cme-s-gamma} \\
\frac{dA_i}{dx} =  \frac{m}{2} k_s A^*_s e^{i\Delta k x}
+ i\frac{3}{8} \gamma k_i A_i ( k_i^2 |A_i|^2 + 2 k_s^2 |A_s|^2).\;\; \label{cme-i-gamma}
\end{eqnarray}
The self-Kerr ($\propto \gamma A_{s,i} |A_{s,i}|^2$) and the cross-Kerr
($\propto \gamma A_{s,i} |A_{i,s}|^2$) nonlinear terms both cause
the effect of phase modulation \cite{Agrawal} and hence phase
mismatch, $\Delta k \neq 0$, \cite{White2015,Macklin2015,Bell-Samolov2015,Zorin2016}
rising with the growth of the signal and the idler.
As result, the total gain of TWJPA is reduced. We solve
Eqs. (\ref{cme-s-gamma}) and (\ref{cme-i-gamma}) with initial conditions
$A_s(0) = A^\textrm{in}_s = \sqrt{2Z_0 P^\textrm{in}_s}/\omega_s \varphi_0$
and $A_i(0) = 0$ numerically and present the results in Fig.~5.

The plot shows almost exponential growth of the signal power
propagating in the array of SQUIDs with effective critical
current $I_0 = 4~\mu$A and driven by the flux wave ensuring the
modulation-parameter value $m = 0.06$.
One can see that for sufficiently small input power the amplifier shows
a nominal gain of about 20~dB (see the bottom curve, corresponding to
$P^\textrm{in}_s \approx -91$~dBm and $P^\textrm{out}_s \approx -71$~dBm).
For somewhat larger input power values (see, e.g., the middle
curve with $P^\textrm{in}_s \approx -88$~dBm
the suppression of the nominal gain is still rather small (i.e., about $-0.2$~dB).
In this case, the amplitude of current oscillations in the output sections
of the signal transmission line approaches
approximately $0.7 I_0$, whereas the Josephson phase oscillations are about $\pi/4$.

The upper solid curve, calculated for the largest input power of
$P_\textrm{1dB} \approx -84$~dBm, exhibits
1-dB suppression of the nominal gain of 20 dB and
hence corresponds to the amplifier compression point.
In this case, the signal current amplitude in the output sections of the line
approaches approximately $0.9 I_0$ leading to an amplitude of the Josephson phase
oscillations of about 1.1 rad.
At such large amplitude the cubic approximation of the Josephson
nonlinearity in the original wave equation (\ref{Eq-motion-signal})
is no longer strictly valid, so the above-mentioned value of $-84$ dBm can be
considered only as an estimate of the gain compression point.
Still, this value is broadly larger than that achievable in
nonlinearity-based TWJPAs with almost similar electrical
parameters (both with four-wave mixing and three-wave mixing).
For example, the gain compression points reported in Refs. \cite{OBrien2014,Zorin2016}
are $-98$ dBm (for $I_0 = 3.29~\mu$A) and $-96$ dBm (for $I_0 = 5~\mu$A),
respectively.

The reason for the notable increase of the amplifier dynamic range
is the almost absence of pump depletion. In the above example, the maximum
output signal power is sufficiently small
(i.e., $P^\textrm{out}_s + P^\textrm{out}_i \approx 2 P^\textrm{out}_s
\approx -62~\textrm{dBm} \ll P_p = -53~\textrm{dBm}$)
and is set by saturation
in the last sections of the line, as one can conclude from
comparison of the solid and dashed lines in Fig.~5.
This property allows us to achieve signal gain in the most-efficient way.

\section{Conclusion and prospects}

We develop a concept for a flux-driven TWJPA with
a large dynamic range. Due to the separate transmission-line, fed through an
isolated input port by sufficiently large pump power, the amplifier is
practically free of the pump-depletion effect.
Moreover, the linear regime of its operation ensures small
distorsions of signal, which prevents the generation of shock waves. \cite{Landauer1960}
Because of good phase matching in a rather broad bandwidth, these properties allow
parametric amplifiers to be designed with rather large number of elementary
cells $N$ and therefore larger gain (e.g., 40~dB).
In combination with ultimately quantum-limited performance over a wide
frequency range, such a TWJPA could be an indispensable device for amplifying
weak signals from quantum sources, including qubits.

When required, even larger output signals can be achieved by using
in each elementary cell not of a single SQUID but of a group of several
serially connected SQUIDs (as reported in Ref.~\cite{Zhou2014}) designed
with Josephson junctions having larger critical
currents and therefore weaker (unwanted) Kerr nonlinearity.
In this case, the need for a cold HEMT preamplifier with typical noise figure
of several kelvins, which is usually unavoidable in
high-fidelity microwave measurements, is eliminated. The latter feature may be especially
useful in quantum-information processing and, in particular, in quantum-interference
and quantum-correlation measurements with single microwave photons. \cite{Bozyigit2010,Peng2016}

The proposed circuit may also operate in the quantum regime enabling efficient
production of entangled photon pairs. As recently demonstrated
by Lahteenm\"{a}ki et al., \cite{Lahteenmaki2013}
a similar array of 250 dc SQUIDs, embedded in a microwave resonator and pumped by
homogeneous alternating flux at the double resonant frequency
of $2\omega_\textrm{res} = 2\pi \times 10.8$~GHz, gave rise to the generation of biphotons
at frequencies $\omega_a$ and $\omega_b$.
The corresponding photon conversion, $\hbar \omega_p =\hbar \omega_a + \hbar \omega_b$,
was, however, possible only in a rather narrow bandwidth of about 200~MHz around
$\omega_\textrm{res} \approx 0.5\omega_p \approx \omega_a \approx \omega_b$.
In this experiment, periodic modulation of the refractive index of
the SQUID metamaterial embedded in the cavity
ensured the photon conversion similar to that occurring in the case of periodic modulation
of the resonator boundary impedance (the mirror position) in the microwave-resonator
version of the dynamical Casimir effect. \cite{Wilson2011}
In our case of pumping by means of a traveling flux wave, the parametric
down-conversion should produce biphotons in a broad frequency range; that is,
like in the recent experiment on production of Casimir photons due to time-varying
boundary impedance of a semi-infinite transmission line. \cite{Schneider2018}
The generated entangled pairs of photons can be used in quantum sensing circuits,
quantum cryptography and other fields of quantum-information science.

Finally, the design of our circuit could
be useful for implementation of the superconducting-circuit analog \cite{Nation2009} of
the event horizon and emission of Hawking radiation. \cite{Hawking1974,Nation2012}
The steplike pulse propagating in the separate transmission line instead of continuous
pump wave can induce a locally decreased speed of light in the SQUID array
with location of the horizon where the propagation velocities in the two
transmission lines coincide. As proposed by Nation et al.,\cite{Nation2009} at sufficient
steepness of the flux steplike pulse (corresponding to sufficiently high Hawking temperature
ensuring visibility of nonclassical radiation above the thermal radiation background)
the circuit should emit detectable (see a possible measuring strategy
in, e.g., Ref. \cite{Besse2018}) microwave photons in the two-mode squeezed state.

\begin{acknowledgments}

The author thanks D. Vion for stimulating discussions and useful
comments concerning the problem of phase matching,
R. Dolata, D. Esteve, M. Khabipov, C. Ki{\ss}ling, Y. Pashkin, A. Ustinov, and M. Weides for their discussions
on possible design of the circuit, and T. Dixon, E. Enrico,  L. Fasolo, and C. Shelly
for their comments on the circuit model.
This work was partially funded by the Joint Research Project PARAWAVE of the
European Metrology Programme for Innovation and Research (EMPIR).
This project received funding from the EMPIR programme co-financed
by the participating states and from the European Union's Horizon 2020
research and innovation programme.

\end{acknowledgments}

\appendix

\section{Equation of motion}

\subsection{The circuit Lagrangian}

Our circuit consists of a linear $LC$ transmission pump line
formed by $N$ identical elements $L'$ and $C'$ and
a signal transmission line formed by the identical symmetric two-junction SQUIDs
and the identical ground capacitances $C$ (see Fig.~1a).
The critical currents and capacitances of the Josephson junctions are nominally
identical, $I_{c1} = I_{c2} = I_0$ and $C_{J1} = C_{J2} = 0.5 C_{J}$ (see Fig.~1b).
We assume that the geometrical
inductances of the SQUID branches $L_{g1}$ and $L_{g2}$ are much smaller than the Josephson
inductance of the individual junctions
($L_{J0} = \varphi_0/I_0$), and are therefore omitted.
Such a SQUID is equivalent to a single junction with a flux-dependent critical
current (see, e.g., Ref. \cite{Likharev1986}),
\begin{equation}
I_c(\Phi_e) = 2I_0 |\cos(\Phi_e /2 \varphi_0)|,
\label{SQUID-Ic}
\end{equation}
and therefore with a flux-dependent potential energy,
\begin{equation}
U_{\textrm{SQUID}}(\Phi_e, \varphi) = - 2E_{J0} |\cos(\Phi_e /2 \varphi_0)| \cos\varphi,
\label{SQUID-U}
\end{equation}
where $E_{J0} = \varphi_0 I_0$ is the Josephson characteristic energy,
$\Phi_e = \Phi_{\textrm{dc}} +\Phi_{\textrm{ac}}$ is the total
external flux applied to the SQUID loop and $\varphi$ is the phase difference on the SQUID.
The kinetic energy of each SQUID is associated with the charging
energies of two junctions, i.e.,
\begin{equation}
K_{\textrm{SQUID}} = \frac{(C_{J1} + C_{J2}) V^2}{2} = \frac{C_J (\varphi_0 \dot{\varphi})^2}{2},
\label{SQUID-K}
\end{equation}
where $\varphi_0 \dot{\varphi}=V$ is the voltage on the SQUID, which has total
capacitance $C_J$ (see the equivalent
electrical circuit in Fig.~1c).

To derive the equations of motion of our electrical circuit with a large number
of variables, we apply Lagrangian mechanics (a similar approach was used, for example,
by Wallquist et al.  \cite{Wallquist2006}) and start by
constructing the Lagrangian $\mathcal{L}$ that describes the entire circuit:
\begin{equation}
\mathcal{L} = \mathcal{L}_p + \mathcal{L}_s,
\label{Lagr1}
\end{equation}
where $\mathcal{L}_p$ and $\mathcal{L}_s$ are the Lagrangians of the
pump line and the signal line, respectively.
In terms of the flux variables associated with the node values $F_n$ and
$\Phi_n$ for the pump and signal lines, respectively, 
these Lagrangians read
\begin{equation}
\mathcal{L}_p = \sum_{n=1}^{N} \frac{C'\dot{F}_n^2}{2} -\frac{(F_{n+1}-F_n)^2}{2L'}
\label{Lagr-p}
\end{equation}
and (see, e.g., the Lagrangian approach to the traveling-wave
parametric amplifier with Kerr nonlinearity in Ref. \cite{Kochetov2016})
\begin{eqnarray}
\mathcal{L}_s = \sum_{n=1}^{N} \frac{C\dot{\Phi}_n^2}{2} + \frac{C_J (\dot{\Phi}_{n+1}-\dot{\Phi}_n)^2}{2} \nonumber ~~~~~~ \\
+ 2E_{J0} \left|\cos \frac{\Phi_{\textrm{dc}} + \kappa (F_{n+1} - F_n)}{2\varphi_0} \right| \cos\frac{\Phi_{n+1}-\Phi_n}{\varphi_0}.
\label{Lagr-s}
\end{eqnarray}
Here, the time derivatives $\dot{F}_n$ and $\dot{\Phi}_n$ are equal to the voltages on the $n$th nodes of the pump
and signal transmission lines, respectively.
The phase difference on the $n$th SQUID is expressed
in terms of magnetic
fluxes on the corresponding nodes, i.e., $\varphi_n =(\Phi_{n+1}-\Phi_n)/\varphi_0$.
The dimensionless coefficient $\kappa = M/L'$ determines the strength of coupling
of the pump and signal lines.

\subsection{Pump transmission line}

In the case of small coupling, $\kappa \ll 1$, compensated by sufficiently large
input pump power,
\begin{equation}
P^{\textrm{in}}_p \approx P^{\textrm{out}}_p  \gg P^{\textrm{out}}_s +P^{\textrm{out}}_i,
\label{Pump-is-large}
\end{equation}
where $P^{\textrm{out}}_{s}$ is the output signal power and $P^{\textrm{out}}_{i}$
is the output idler power, and
the backaction of the signal line on the pump line can be ignored.
Then the decoupled equation of motion
for fluxes $F_n$ can be obtained from the Euler-Lagrange equation
\begin{equation}
\frac{d}{dt} \frac{\partial \mathcal{L}}{\partial \dot{F}_n}
- \frac{\partial \mathcal{L}}{\partial F_n}\approx\frac{d}{dt} \frac{\partial \mathcal{L}_p}
{\partial \dot{F_n}} - \frac{\partial \mathcal{L}_p}{\partial F_n}= 0.
\label{Eul-Lagr-p}
\end{equation}
With use of Eq. (\ref{Lagr-p}), this equation reads
\begin{equation}
\ddot{F}_n + \omega'^2_0(-F_{n-1} + 2 F_n - F_{n+1})= 0,
\label{Eq-F}
\end{equation}
where $\omega'_0 = 1/\sqrt{L'C'}$ is the cutoff frequency of the pump transmission line.
Here, the index range is $n = 1,2,...,N-1,$
and the boundary values are $F_0 = F_{\textrm{in}}$ and $F_{N} = F_{\textrm{out}}$.

The set of coupled differential equations given by (\ref{Eq-F})
is the discrete
analog of the partial differential equation describing plane waves.
It is easy to verify by substitution that, for a sufficiently small
frequency $\omega_p \ll \omega'_0$ (i.e., when the wavelength is larger
than the size of the elementary cell, $\lambda_p \gg a$), the solution describing
the wave traveling, for example, in the right direction has the shape
\begin{equation}
F_n \propto e^ {i(k_p n - \omega_p t + \chi_p)},
\label{Sol-F}
\end{equation}
where $k_p$ is the wave number normalized on the reverse size of the cell $a^{-1}$,
and $\chi_p$ is an initial phase.
Incorporation of Eq.~(\ref{Sol-F}) into Eq.~(\ref{Eq-F}) yields
\begin{equation}
-\left[ \omega^2_p -\omega'^2_0 (-e^{-ik_p} + 2 - e^{ik_p})\right]F_n = 0
\label{Eq-subst}
\end{equation}
or, equivalently,
\begin{equation}
\left[\omega^2_p - 4\omega'^2_0 \sin^2(k_p/2)\right]F_n = 0.
\label{Eq-subst2}
\end{equation}
This equation determines the dispersion relation of the transmission line
(see, for example, Ref.~\cite{Martin2015}),
\begin{equation}
k_p(\omega_p) = \pm 2 \arcsin \left( \omega_p/2\omega'_0 \right),
\label{disp-kp}
\end{equation}
which in our case of positive small $k_p$ yields the relation
\begin{equation}
k_p \approx \frac{\omega_p}{\omega'_0} \left(1 + \frac{\omega^2_p}{24\omega'^2_0}  \right).
\label{disp-kp2}
\end{equation}
This formula describes the standard positive dispersion, $d^2k_p/d\omega_p^2 > 0$.

Without loss of generality, we set the pump phase $\chi_p$ in Eq.~(\ref{Sol-F}) in such a way
that the pump wave has the form
\begin{equation}
F_n = \varphi_0 A_{p0} \cos[k_p (n-0.5) - \omega_p t],
\label{Sol-F2}
\end{equation}
where dimensionless amplitude $A_{p0}$ is real and positive.
In this case, current $I^{(n)}_p$ flowing through inductance $L'$ in the $n$th cell is expressed as
\begin{eqnarray}
I^{(n)}_p = -(F_{n+1}- F_n)/L'
\label{current-Ip}
\end{eqnarray}
and the flux induced in the $n$th cell of the signal transmission line is
\begin{eqnarray}
M I^{(n)}_p &=& -\kappa (F_{n+1}- F_n) \nonumber \\
&=& 2 \varphi_0 \kappa A_{p0} \sin\frac{k_p}{2} \sin(k_p n - \omega_p t) \nonumber \\
&\approx& \kappa \varphi_0 A_{p0} k_p \sin(k_p n - \omega_p t).
\label{inducedFlux}
\end{eqnarray}
This formula describes the wave of magnetic flux that ensures the necessary time variation
of the SQUID inductance.

\subsection{Signal transmission line}

Incorporating Eq. (\ref{inducedFlux}) into Eq.~(\ref{Lagr-s}), we obtain
in the adiabatic approximation the equations of motion for the fluxes $\Phi_n$,
\begin{equation}
\frac{d}{dt} \frac{\partial \mathcal{L}}{\partial \dot{\Phi}_n} - \frac{\partial \mathcal{L}}{\partial \Phi_n}=
\frac{d}{dt} \frac{\partial \mathcal{L}_s}
{\partial \dot{\Phi_n}} - \frac{\partial \mathcal{L}_s}{\partial \Phi_n}= 0,
\label{Eul-Lagr-s}
\end{equation}
or
\begin{eqnarray}
C\ddot{\Phi}_n &+& C_J (-\ddot{\Phi}_{n-1}+2\ddot{\Phi}_n - \ddot{\Phi}_{n+1}) \nonumber \\
&+& I^{(n)}_c(t) \sin[(\Phi_{n}-\Phi_{n-1})/\varphi_0]\nonumber \\  
&-& I^{(n+1)}_c(t) \sin[(\Phi_{n+1}-\Phi_n)/\varphi_0] = 0.    
\label{Eq-Phi-n}
\end{eqnarray}
Here the time-dependent critical current of the $n$th SQUID is
\begin{equation}
I^{(n)}_c(t) = I_{c0}[1 + m \sin(k_p n - \omega_p t)],
\label{Ic-n}
\end{equation}
where magnitude $I_{c0}$ is determined by the constant flux $\Phi_{\textrm{dc}}$; that is,
\begin{equation}
I_{c0} = 2I_0 |\cos(\Phi_{\textrm{dc}} /2 \varphi_0)|.
\label{Ic0}
\end{equation}
The small, positive value of $m \ll 1$ is found by linearizing Eq.~(\ref{SQUID-Ic})
in vicinity of the optimal working point $\Phi_\textrm{dc} = \pi\varphi_0/3$ (see Fig.~2),
\begin{equation}
m = \frac{\kappa}{2}k_p A_{p0} |\tan (\Phi_{\textrm{dc}} /2 \varphi_0)| \approx \frac{\sqrt{3}}{2}\kappa k_p A_{p0}.
\label{m-value}
\end{equation}
Assuming that the phases on the SQUIDs are small, i.e., $|\varphi_{n}| \ll 1$,
we make the approximation
\begin{equation}
\sin \varphi_{n} \approx \varphi_{n} - \varphi_{n}^3/6,
\label{sin-expans}
\end{equation}
and ignore high-order cross terms $\propto m \varphi_{n}^3$. We thus replace the $n$th SQUID
with the time-dependent (in the general case, slightly nonlinear) inverse inductance
$1/L^{(n)}_{J}(t,\varphi_{n})$; that is,
\begin{equation}
\frac{L_{J0}}{L^{(n)}_{J}} = 1 + m \sin(k_p n - \omega_p t) - \frac{1}{2\varphi_0^2}(\Phi_{n+1}-\Phi_{n})^2,
\label{SQUID-ind}
\end{equation}
where $L_{J0} = \varphi_0/I_0$.
Incorporating Eq.~(\ref{Ic-n}) into Eq.~(\ref{Eq-Phi-n})
and using approximation (\ref{sin-expans}), we obtain
\begin{eqnarray}
\omega^{-2}_0 \ddot{\Phi}_n -  \omega^{-2}_J (-\ddot{\Phi}_{n-1}+2\ddot{\Phi}_n - \ddot{\Phi}_{n+1}) \nonumber \\
+ [1+m \sin(k_p n - \omega_p t)] (\Phi_n-\Phi_{n-1}) \nonumber \\
- \{1+m \sin[k_p (n+1) - \omega_p t]\} (\Phi_{n+1}-\Phi_n) \nonumber \\
- \frac{1}{6\varphi_0^{2}} \left[(\Phi_n-\Phi_{n-1})^3- (\Phi_{n+1}-\Phi_n)^3\right] = 0.  
\label{Eq-Phi-n-2}
\end{eqnarray}
Here, the cutoff frequency of the bare transmission line $\omega_0$ is $1/\sqrt{L_{J0} C}$
and the Josephson plasma
frequency $\omega_J$ is $1/\sqrt{L_{J0} C_J}$.

For sufficiently small amplitude, $|\varphi_{n}| \ll 1$, small
frequency, $\omega \ll \omega_0, \omega_J$, and the absence
of a pump ($m = 0$), a solution of Eq.~(\ref{Eq-Phi-n-2}) has the form of the plane wave
\begin{equation}
\Phi_n \propto e^ {i(k n - \omega t + \chi)},
\label{Sol-Phi}
\end{equation}
where $k$ is the wave number and $\chi$ is the phase.
Incorporation of this expression into Eq.~(\ref{Eq-Phi-n-2}) yields
\begin{equation}
-\left[ \omega^2 + \omega^2_0 \left(\frac{\omega^2}{\omega^2_J} -1 \right)(-e^{-ik} + 2 - e^{ik})\right]\Phi_n = 0.
\label{Eq-subst22}
\end{equation}
The corresponding dispersion equation reads
\begin{equation}
\omega^2 - 4\omega^2_0 \left(1 - \frac{\omega^2}{\omega^2_J}\right)\sin^2(k/2) = 0
\label{Eq-subst3}
\end{equation}
and has the solution
\begin{equation}
k(\omega) = \pm 2 \arcsin \left( \frac{\omega/2\omega_0}{\sqrt{1-\omega^2/\omega^2_J}} \right) = 0,
\label{disp-k}
\end{equation}
which for the wave propagating in the right direction ($k>0$) can
be approximated as
\begin{equation}
k(\omega) \approx \frac{\omega}{\omega_0} \left( 1 + \frac{\omega^2}{2\omega^2_J} + \frac{\omega^2}{24\omega^2_0} \right).
\label{disp-k-approx}
\end{equation}
In comparison with Eq.~(\ref{disp-kp2}), this dispersion relation
includes the term $\omega^2/2\omega^2_J$,
which stems from the Josephson plasma resonance in the SQUIDs.

\subsection{Continuum approximation}

In the case of small frequency $\omega$ (and therefore, a large wavelength $\lambda
= 2\pi/k \approx 2\pi \omega_0/\omega \gg 1$),
the equation of motion (\ref{Eq-Phi-n}) can be presented in terms of partial derivatives
of continuous variables (see, for example, Refs. \cite{Yaakobi2013} and~ \cite{Kochetov2016}).
We introduce the flux variable $\Phi(x,t)$,
whose values on the grid $x = n$ with unity step, $\Delta x = 1$, coincide with $\Phi_n(t)$,
\begin{equation}
\Phi_n(t) \rightarrow \Phi(x,t).
\label{cont-phi}
\end{equation}
Thus, variable $x$ is a dimensionless coordinate, whereas the genuine coordinate variable
is $X = ax$, where $a$ is the cell size \cite{Zorin2016}.
The parameter-modulation function
$f_n(t) = [1+m \sin(k_p n - \omega_p t)]$ in Eq.~(\ref{Eq-Phi-n-2}) is now replaced by
a continuous function; that is,
\begin{equation}
f_n(t) \rightarrow f(x,t) = [1+m \sin(k_p x - \omega_p t)].
\label{mod-func}
\end{equation}
Following the method derived by Yaakobi et al.,\cite{Yaakobi2013} the finite differences
can be expressed by partial derivatives of
a continuous function $\mathcal{F}(x)$ according to the following rules:
\begin{eqnarray}
\mathcal{F}_{n+1}-\mathcal{F}_n = \frac{\partial \mathcal{F}}{\partial x}, \label{deriv-approx1} \\
\mathcal{F}_{n+1}-2\mathcal{F}_n + \mathcal{F}_{n-1} = \frac{\partial^2 \mathcal{F}}{\partial x^2}.
\label{deriv-approx}
\end{eqnarray}

Then, the set of the finite-difference equations (\ref{Eq-Phi-n}) takes the form of
a continuum wave equation,
\begin{eqnarray}
 \frac{\partial^2 \Phi}{\partial x^2} - \omega^{-2}_0\frac{\partial^2 \Phi}{\partial t^2}
+ \omega^{-2}_J \frac{\partial^4 \Phi}{\partial t^2\partial x^2} - \frac{1}{6\varphi_0^{2}} \frac{\partial}{\partial x}
\left[ \left( \frac{\partial \Phi}{\partial x} \right)^3 \right] \nonumber \\
+ \frac{\partial}{\partial x} \left[ m \sin(k_p x - \omega_p t) \frac{\partial \Phi}{\partial x} \right] = 0,~~~
\label{Eq-Phi-n7}
\end{eqnarray}
including the wave-like variation of the distributed linear inductance
of the transmission line,
\begin{equation}
L^{(n)}_{J}(t) \rightarrow L_{J}(x,t) = \frac{L_{J0}}{1 + m \sin(k_p x - \omega_p t)}.
\label{SQUID-distr-ind}
\end{equation}
The fourth term on the left-hand-side of Eq.~(\ref{Eq-Phi-n7}) describes
nonlinear effects and it is essential only
in cells where the amplitude of the signal or idler is not sufficiently small.

\section{Effect of parameter variations}

Sufficiently small variations of the line parameters is the basic requirement
for proper operation of Josephson parametric amplifies based on traveling microwaves
(e.g., see the evaluation of fabrication tolerances in
$LC$ cells enabling resonant phase matching in a four-wave-mixing TWJPA in Ref.~\cite{OBrien2014}).
As long as in our circuit the pump and the signal (idler) waves travel along two different
lines, the appreciable parameter variations in either line may cause significant
phase mismatch. The signal line including the Josephson junctions and thereby
having a more-complex design, is therefore more prone to such variations.

\subsection{Inaccuracy of magnetic flux setting}

The condition of zero phase mismatch for a small detuning, $|\delta| \ll 1$,
is given by Eqs.~(\ref{dk0}), (\ref{eta-def}),
and (\ref{nu-def}), i.e.,
\begin{eqnarray}
\Delta k &=& k_p - k_i - k_s = \nu - \eta/3 \nonumber \\
&=& \frac{\omega_p}{\omega'_0} -\frac{\omega_p}{\omega_0}
 \left[1+\frac{\omega^2_p}{8\omega^2_J} \right] = 0. \label{dk000}
\end{eqnarray}
Because the cutoff frequency of the bare transmission line, $\omega_0$,
and the plasma frequency, $\omega_J$, are
both proportional to $I_0^{1/2}(\Phi_\textrm{dc})$, this condition can be fulfilled
by applying the optimal dc flux bias, $\Phi_\textrm{dc} =\Phi^\textrm{opt}_\textrm{dc}$.
In the case of inaccurate flux setting or instability of the flux bias in time,
$\Phi_\textrm{dc} =\Phi^\textrm{opt}_\textrm{dc} + \delta\Phi_\textrm{dc}$,
the corresponding deviation of the Josephson critical current from
the optimal value $I_0$ is
\begin{equation}
\delta I_0 = \frac{\partial I_c}{\partial \Phi_\textrm{dc}} \delta\Phi_\textrm{dc} = \epsilon I_0.
\label{LJ-dev}
\end{equation}
Here we used the notation $\epsilon = - \sqrt{3}~ \delta\Phi_\textrm{dc}/ \varphi_0$
and applied formula ({\ref{Ic0}}) in the point $\Phi_\textrm{dc}/2\varphi_0 = \pi/3$.
The nonzero value $\delta I_0$ causes deviation of the cutoff and plasma frequencies,
$\omega_{0,J} \rightarrow \omega_{0,J} + \delta \omega_{0,J}$, where
\begin{equation}
\frac{\delta \omega_{0,J}}{\omega_{0,J}} = \frac{1}{\omega_{0,J}} \frac{\partial \omega_{0,J}}{\partial I_0} \delta I_0
= 0.5 \epsilon.
\label{cutoff-f-dev}
\end{equation}
The corresponding phase mismatch is
\begin{eqnarray}
\Delta k  &=& \frac{\omega_p}{\omega'_0} -\frac{\omega_p}{\omega_0 (1 + 0.5 \epsilon)}
 \left[1+\frac{\omega^2_p}{8\omega^2_J (1 + 0.5 \epsilon)^2} \right] \nonumber \\
&\approx& 0.5 \epsilon ~ \omega_p/\omega_0,
\label{mismatch-dF}
\end{eqnarray}
where a small term $\propto \omega^2_p/\omega^2_J$ has been omitted.

Taking the target frequency $\omega_p = 0.2\omega_0$ (yielding $\Delta k = 0.1 \epsilon$),
a modulation coefficient of $m = 0.06$, an exponential
gain coefficient $g_0$ of $ 0.25 m \omega_p/\omega_0 = 0.003$,
and the line length $N=1000$ (yielding a nominal gain of 20~dB), we obtain the corresponding suppression
of the gain (see Eq.~(\ref{general-case-power-gain})):
\begin{equation}
G/G_\textrm{max} = 1 - \left(\frac{\Delta k}{2g_0}\right)^2 \tanh^2 g_0 N \approx 1- 275 ~\epsilon ^2.
\label{G-reduction}
\end{equation}
This formula yields the $1$-dB reduction of the gain
for reasonably small value of $|\epsilon| \approx 0.02$ or, equivalently,
reasonably small inaccuracy in the magnetic flux setting
of $ |\delta\Phi_\textrm{dc}| = \epsilon \varphi_0 /\sqrt{3} \approx 0.01 \varphi_0$.

\subsection{Signal line inhomogeneity}

In practical circuits, the electrical parameters can have a somewhat
irregular distribution over the length of the line.
This primarily concerns the Josephson inductance $L_{J0}(x)$, whose local value
depends on the critical current $I_{0}$ of corresponding SQUID. This
critical current depends on the area of the Josephson junction, the local
critical current density $j_c$ and the offset magnetic flux $\Phi_\textrm{dc}$, whose value,
under the assumption of a homogeneous magnetic field, depends on the size and shape of the SQUID loop.
In fabrication, however, variations of these parameters can be controlled only
within certain limits.

In our circuit, which comprises Josephson junctions with an area
of about 1~$\mu$m$^2$, the parameter which is most prone to random variations is the
critical current. To roughly model small variations of the critical current $\delta I_0(x)$
around its mean value $\langle I_0 \rangle$ (leading to small variations of the reverse
Josephson inductance
$\delta L^{-1}_{J0}$ around the mean value $\langle L^{-1}_{J0} \rangle)$, we should
add a corresponding small random term to the linear part of equation of motion (\ref{Eq-Phi-n7}), i.e.,
\begin{eqnarray}
\frac{\partial^2 \Phi}{\partial x^2} - \omega^{-2}_0\frac{\partial^2 \Phi}{\partial t^2}
+ \omega^{-2}_J \frac{\partial^4 \Phi}{\partial t^2\partial x^2} ~~~~~~~~~~~~~~~~~~~~~~ \nonumber \\
+ \frac{\partial}{\partial x} \left\{ \left[ m \sin(k_p x - \omega_p t) + \zeta(x) \right] \frac{\partial \Phi}{\partial x} \right\} = 0,
\label{Eq-Phi-n8}
\end{eqnarray}
where
\begin{equation}
\zeta(x)  =\delta I_0(x)/\langle I_0 \rangle =\delta L^{-1}_{J0}(x)/\langle L^{-1}_{J0}\rangle
\label{h-x}
\end{equation}
is a random dimensionless function.

This function is defined on the nodes, $x = n$, and has, presumably, a small rms
value, $\sqrt{\langle\zeta^2(x)\rangle} = \sigma  \ll 1$
and a short correlation length ($\ell_c \sim 1$).
Here, the sign $\langle ...\rangle$ denotes
averaging over the statistical ensemble.
The cutoff and plasma frequencies are now
defined as
\begin{equation}
\omega_0 =1/\sqrt{\langle L_{J0} \rangle C}
~~\textrm{and}~~\omega_J = 1/\sqrt{\langle L_{J0}\rangle C_J},
\label{new w}
\end{equation}
respectively, whereas the random values $\zeta(x)$ cause local
variations of these frequencies and therefore of the wave number, which leads
to a phase mismatch.
The small higher-order term $\propto m \zeta(x)$ describing variation of
the parametric coupling is omitted here.

To roughly evaluate the random drift of the phase and therefore
the resulting phase mismatch 
we put in Eq.~(\ref{Eq-Phi-n8}) the zero pump, $m = 0$, and find the
solution in the simple wave form:
\begin{eqnarray}
\Phi(x,t) = \frac{\varphi_0 }{2} A e^{i[k x-\omega t + \Theta(x)]}+\textrm{c.c.}
\label{solution-xi-effect}
\end{eqnarray}
Here, the complex phase $\Theta(x) = \psi (x) - i \rho(x)$ includes the
random phase as such, $\psi (x)$, and the logarithm of the random
amplitude, $\rho(x) = \ln [A(x)/A_0]$.
Both these variables are induced by small perturbation $|\zeta(x)| \ll k$  and slowly vary in
space, i.e. $ \left| \frac{\partial \psi}{\partial x} \right|, \left|
\frac{\partial \rho}{\partial x} \right| \ll k $.
Keeping only essential terms in Eq.~(\ref{Eq-Phi-n8}), we obtain two decoupled equations
for $\psi (x)$ and $\rho(x)$:
\begin{eqnarray}
\frac{d \rho}{dx} &=& \frac{1}{2}\frac{d \zeta(x)}{dx}, \label{mod-coupled-mode-eq1a} \\
\frac{d\psi}{dx} &=&  -\frac{k}{2} \zeta(x).   \label{mod-coupled-mode-eq1b}
\end{eqnarray}
Equation (\ref{mod-coupled-mode-eq1a}) yields the solution
for the wave amplitude, $A(x) = A_0 [1+ 0.5 \zeta(x)]$,
whose space variation mimics the local fluctuation of the transmission line admittance,
\begin{equation}
Y(x) = \sqrt{C/L_{J0}(x)} = \left[1 + 0.5 \zeta(x)\right]Y_0,
\label{local-admittance}
\end{equation}
around its average value $Y_0=1/Z_0$.

The solution of equation (\ref{mod-coupled-mode-eq1b}) for the wave phase $\psi$
has the form
\begin{equation}
\psi(x) = - 0.5 k \int_0^x \zeta(x') dx'.
\label{phases-solution}
\end{equation}
This formula describes the diffusion of phase $\psi(x)$ (the Brownian-motion process) 
and therefore yields the statistical average values $\langle \psi(x) \rangle = 0$
and $\langle \psi^2(x) \rangle = 0.25 k^2 \sigma x$.
The corresponding drift of the phase on the
length $x = N$ can be estimated as
$\Psi = \sqrt{\langle\psi^2(N)\rangle} = 0.5k\sqrt{\sigma N}$.
For critical current variations $\sigma$ on the order of 0.05 (see, e.g., Ref.~ \cite{Nakada2003}),
dimensionless wave number $k =0.1$,
and line length $N = 1000$, this formula yields $\Psi \approx 0.35$~rad and therefore
still negligibly small suppression of the power gain due to such phase mismatch, i.e.,
\begin{equation}
G/G_\textrm{max} = 1 - \left(\frac{\Psi}{2g_0 N}\right)^2 \tanh^2(g_0 N) \approx 0.997.
\label{G-reduction2}
\end{equation}

Another problem that may arise in fabrication is a notable gradient of the local critical
current density $\nabla j_c(\textbf{r})$ of the Nb trilayer, which may account for
a drop of up to 30~\% in $j_c$ over the 3-in. wafer. \cite{Khabipov1018}
In this case, inhomogeneity of the distributed inductance in the
straight transmission line takes the form  $L_{J0}(x) = L^{(0)}_{J0}/(1+\mu x)$.
This situation can also be modeled by Eq.~(\ref{Eq-Phi-n8}) with
the regular function $\zeta(x)$ having the shape $\zeta(x) = 0.5 \mu x$.
For example, for the total length of the line $\ell = a N \sim 5$~cm, the
product $\mu N$ can be on the order of 0.1.
The corresponding maximum phase drift $\Psi = 0.25 k \mu N^2$ can then
approach the excessively large value of 2.5~rad.
This unwanted effect resulting from linear variation of $j_c$ can possibly
be mitigated by applying
the offset magnetic field with a small gradient
in the direction of the transmission line. This field should compensate
the regular change in the critical currents of the junctions via a
corresponding change of offset flux in the SQUIDs. Although this
method leads to a more-complicated experiment, it may seemingly also work in the case
when the SQUID array has a meander shape.

Finally, appreciable inhomogeneities of the circuit parameters with correlation length
comparable to characteristic wavelengths (in our case, about 30-60 elementary cells)
may cause partial reflections of traveling waves inside the transmission lines and hence
deteriorate the TWJPA performance. Analysis of such a case, which may also occur
in conventional TWJPAs based on Josephson nonlinearity, is, however,
beyond the scope of this work.


\begin{thebibliography}{26}


\bibitem{Castellanos-Beltran2008} M. A. Castellanos-Beltran, K. Irwin, G. Hilton, L. Vale, and K. Lehnert,
 Nat. Phys. {\bf 4}, 928 (2008).

\bibitem{Clerk2010} A. A. Clerk, M. H. Devoret, S. M. Girvin, F. Marquardt, and R. J. Schoelkopf,
Rev. Mod. Phys. {\bf 82}, 1155  (2010).

\bibitem{Abdo2011} B. Abdo, F. Schackert, M. Hatridge, C. Rigetti, and M. Devoret,
Appl. Phys. Lett. {\bf 99}, 162506 (2011).

\bibitem{Vijay2011} R. Vijay, D. H. Slichter, and I. Siddiqi,
Phys. Rev. Lett. {\bf 106}, 110502 (2011).

\bibitem{Flurin2012} E. Flurin, N. Roch, F. Mallet, M. H. Devoret, and B. Huard, Phys. Rev.
Lett. {\bf 109}, 183901 (2012).

\bibitem{Devoret2013} M.H. Devoret and R.J. Schoelkopf, Science {\bf 339}, 1169 (2013).


\bibitem{Eichler2014} C. Eichler, Y. Salathe, J. Mlynek, S. Schmidt, and A. Wallraff,
Phys. Rev. Lett. {\bf 113}, 110502 (2014).

\bibitem{Vool2016} U. Vool, S. Shankar, S. O. Mundhada, N. Ofek, A. Narla, K. Sliwa,
E. Zalys-Geller, Y. Liu, L. Frunzio, R. J. Schoelkopf, S. M. Girvin, and M. H. Devoret,
Phys. Rev. Lett. {\bf 117}, 133601 (2016).


\bibitem{Lecocq2017} F. Lecocq, L. Ranzani, G. A. Peterson, K. Cicak, R. W. Simmonds,
J. D. Teufel, and J. Aumentado, Phys. Rev. Applied {\bf 7}, 024028 (2017).

\bibitem{Sivak2019} V.V. Sivak, N.E. Frattini, V.R. Joshi, A. Lingenfelter, S. Shankar, and M.H. Devoret,
Phys. Rev. Applied {\bf 11}, 054060 (2019).


\bibitem{Mohebbi2011} H. R. Mohebbi, 
PhD thesis (University of Waterloo, Ontario, Canada, 2011).

\bibitem{Yaakobi2013} O. Yaakobi, L. Friedland, C. Macklin, and I. Siddiqi,
Phys. Rev. B {\bf 87}, 144301 (2013).

\bibitem{OBrien2014} K. O'Brien, C. Macklin, I. Siddiqi, and X. Zhang, Phys. Rev. Lett. {\bf 113}, 157001 (2014).

\bibitem{White2015} T. C. White, J. Y. Mutus, I.-C. Hoi, R. Barends, B. Campbell, Yu Chen, Z. Chen, B. Chiaro, A. Dunsworth,
E. Jeffrey, J. Kelly, A. Megrant, C. Neill, P. J. J. O'Malley, P. Roushan, D. Sank, A. Vainsencher, J. Wenner, S. Chaudhuri,
J. Gao and J. M. Martinis, Appl. Phys. Lett. {\bf 106}, 242601 (2015).

\bibitem{Macklin2015} C. Macklin, K. O'Brien, D. Hover, M. E. Schwartz, V. Bolkhovsky,
X. Zhang, W. D. Oliver, I. Siddiqi, Science {\bf 350}, 307 (2015).

\bibitem{Bell-Samolov2015} M. T. Bell and A. Samolov, Phys. Rev. Applied {\bf 4}, 024014 (2015).

\bibitem{Zorin2016} A. B. Zorin, Phys. Rev. Applied {\bf 6}, 034006 (2016).

\bibitem{Zorin2017}	A. B. Zorin, M. Khabipov, J. Dietel, and R. Dolata, ArXiv:1705.02859.

\bibitem{Miano2018} A. Miano and O. A. Mukhanov, IEEE Trans.
Appl. Supercond. {\bf 29}, 1501706 (2019);  ArXiv:1811.02703.

\bibitem{WenyuanZhang2017} W. Zhang, W. Huang, M. E. Gershenson, and M.~T. Bell,
Phys. Rev. Applied {\bf 8}, 051001 (2017).

\bibitem{Frattini2017} N. E. Frattini, U. Vool, S. Shankar, A. Narla, K. M. Sliwa,
and M. H. Devoret, Appl. Phys. Lett. {\bf 110}, 222603 (2017).

\bibitem{Frattini2018} N. E. Frattini, V. V. Sivak, A. Lingenfelter, S. Shankar, and M. H. Devoret,
Phys. Rev. Applied {\bf 10}, 054020 (2018).

\bibitem{Agrawal} G. P. Agrawal, \emph{Nonlinear fiber optics} (Academic press, San Diego, California, 2007).

\bibitem{Eom2012} B. H. Eom, P. K. Day, H. G. LeDuc, and J. Zmuidzinas,
Nat. Phys. {\bf8}, 623 (2012).

\bibitem{Kamal2009} A. Kamal, A. Marblestone, and M. Devoret,
Phys. Rev. B {\bf79}, 184301 (2009).

\bibitem{Abdo2013} B. Abdo, A. Kamal, and M. Devoret,
Phys. Rev. B {\bf87}, 014508 (2013).

\bibitem{Kylemark2006} P. Kylemark, H. Sunnerud, M. Karlsson, and P. A. Andrekson,
J. Lightwave Techn. {\bf 24}, 3471 (2006).

\bibitem{Landau-Lifshitz-1}  L. D. Landau and E. M. Lifshitz, \emph{Mechanics:
volume 1 of a course of theoretical physics} (Pergamon Press, Oxford, 1969).

\bibitem{Migulin} V.~Migulin, V.~Medvedev, E.~Mustel, and V.~Parygin,
\emph{Basic theory of oscillations} (Mir Publishers, Moscow, 1983).

\bibitem{Yamamoto2008} T. Yamamoto, K. Inomata, M. Watanabe, K. Matsuba, T. Miyazaki, W. D. Oliver,
Y. Nakamura, and J. S. Tsai, Appl. Phys. Lett. {\bf93}, 042510 (2008).

\bibitem{Wilson2010} C. M. Wilson, T. Duty, M. Sandberg, F. Persson, V. Shumeiko, and P. Delsing,
Phys. Rev. Lett. {\bf 105}, 233907 (2010).

\bibitem{Wilson2011} C. M. Wilson, G. Johansson, A. Pourkabirian, M. Simoen,
J. R. Johansson, T. Duty, F. Nori, and P. Delsing, Nature {\bf 479}, 376 (2011).

\bibitem{Lin2013} Z. R. Lin, K. Inomata, W. D. Oliver, K. Koshino, Y. Nakamura, J. S. Tsai,
and T. Yamamoto, Appl. Phys. Lett. {\bf103}, 132602 (2013).


\bibitem{Simbierowicz2018} S. Simbierowicz, V. Vesterinen, L. Gr\"{o}nberg, J Lehtinen, M. Prunnila, and J. Hassel,
Supercond. Sci. Technol. {\bf31}, 105001 (2018).


\bibitem{Zhong2013} L. Zhong, E. P. Menzel, R. Di Candia, P. Eder, M. Ihmig, A. Baust, M. Haeberlein, E. Hoffmann,
K. Inomata, T. Yamamoto, Y. Nakamura, E. Solano, F. Deppe, A. Marx, and R. Gross,
New J. Phys. {\bf15}, 125013 (2013).


\bibitem{Krantz2013} P. Krantz, Y. Reshitnyk, W. Wustmann, J. Bylander, S. Gustavsson,
W. D. Oliver, T. Duty, V. Shumeiko, and P. Delsing,
New J. Phys. {\bf15}, 105002 (2013).


\bibitem{Boutin2017} S. Boutin, D. M. Toyli, A. V. Venkatramani, A. W. Eddins, I. Siddiqi, and A. Blais,
Phys. Rev. Applied {\bf8}, 054030 (2017).

\bibitem{Zhou2014} X. Zhou, V. Schmitt, P. Bertet, D. Vion, W. Wustmann, V. Shumeiko, and D. Esteve,
Phys. Rev. B {\bf89}, 214517 (2014).


\bibitem{Tien1958} P. K. Tien, J. Appl. Phys. {\bf 29}, 1347 (1958).

\bibitem{Cullen1960} A. L. Cullen, Proc. IEE - Part B: Electron. and Communication Eng. {\bf 107}, 101 (1960).

\bibitem{Boyd2008} R. W.~Boyd, {\it Nonlinear optics} (Academic Press, London, 2008).

\bibitem{Erickson2017} R. P. Erickson and D. P. Pappas,
Phys. Rev. B {\bf 95}, 104506 (2017).

\bibitem{Manley-Rowe1956} J. M. Manley and H. E. Rowe,
Proc. IRE {\bf44}, 904 (1956).

\bibitem{Yariv1967} A. Yariv, {\it Quantum electronics} (Wiley, New York, 1967).

\bibitem{Dolata2005} R. Dolata, H. Scherer, A. B. Zorin, and J. Niemeyer, J. Appl. Phys. {\bf 97}, 054501 (2005).

\bibitem{Tolpygo2017} S. K. Tolpygo, V. Bolkhovsky, S. Zarr, T. J. Weir, A. Wynn, A. L. Day,
L. M. Johnson, and M. A. Gouker,  IEEE Trans. Appl. Supercond. {\bf 27}, 1100815 (2017).

\bibitem{Niemeyer1974} J. Niemeyer, PTB-Mitt. {\bf 84}, 251 (1974); G. J. Dolan, Appl. Phys. Lett. {\bf 31},
337 (1977).

\bibitem{van-der-Zant1994} H. S. J. van der Zant, R. A. M. Receveur,
T. P. Orlando, and A. W. Kleinsasser, Appl. Phys. Lett. {\bf 65}, 2102 (1994).

\bibitem{Landauer1960} R. Landauer, IBM J. Res. $\&$ Develop. {\bf 4}, 391 (1960).

\bibitem{Bozyigit2010} D. Bozyigit, C. Lang, L. Steffen, J. M. Fink, C. Eichler, M. Baur,
R. Bianchetti1, P. J. Leek, S. Filipp, M. P. da Silva, A. Blais, and A.Wallraff, Nat. Phys. {\bf7}, 154 (2010).

\bibitem{Peng2016} Z. H. Peng, S. E. de Graaf, J. S. Tsai, and O. V. Astafiev, Nat. Commun. {\bf 7}, 12588 (2016).

\bibitem{Lahteenmaki2013} P. Lahteenm\"{a}ki, G. S. Paraoanu, J. Hassel and P. J. Hakonen,
PNAS {\bf 110}, 4234 (2013).


\bibitem{Schneider2018} B. H. Schneider, A. Bengtsson, I.M. Svensson, T. Aref, G. Johansson,
J. Bylander, and P. Delsing, arXiv:1802.05529.

\bibitem{Nation2009} P. D. Nation, M. P. Blencowe, A. J. Rimberg, and E. Buks,
Phys. Rev. Lett. {\bf 103}, 087004 (2009).


\bibitem{Hawking1974} S. W. Hawking, Nature {\bf 248}, 30 (1974).

\bibitem{Nation2012} P. D. Nation, J. R. Johansson, M. P. Blencowe, and F. Nori,
Rev. Mod. Phys. {\bf 84}, 1 (2012).


\bibitem{Besse2018} J.-C. Besse, S. Gasparinetti, M. C. Collodo,
T. Walter, P. Kurpiers, M. Pechal, C. Eichler, and A. Wallraff,
Phys. Rev. X {\bf 8}, 021003 (2018).


\bibitem{Likharev1986} K. K.~Likharev, {\it Dynamics of Josephson junctions and
circuits} (Gordon and Breach, New York, 1986).

\bibitem{Wallquist2006} M. Wallquist, V. S. Shumeiko, and G. Wendin, Phys. Rev. B {\bf 74}, 224506 (2006).

\bibitem{Kochetov2016} B. A. Kochetov and A. Fedorov, 
2016 8th International Conference
on Ultrawideband and Ultrashort Impulse Signals (UWBUSIS), 5-11 Sept. 2016, Odessa,
Ukraine. IEEE Xplore 112 (2016).

\bibitem{Martin2015} F. Mart\'{\i}n, {\it Artificial transmission lines
for rf and microwave applications},
Wiley Series in Microwave and Optical Engineering (Wiley, 2015).


\bibitem{Nakada2003} D. Nakada, K. K. Berggren, E. Macedo, V. Liberman,
and T. P. Orlando, IEEE Trans. Appl. Supercond. {\bf 13}, 111 (2003).

\bibitem{Khabipov1018} M. Khabipov, private communication.


\end{thebibliography}
\end{document}